\documentclass[a4paper,onecolumn,11pt, unpublished]{quantumarticle}
\pdfoutput=1
\usepackage{braket}
\usepackage[utf8]{inputenc}
\usepackage[english]{babel}
\usepackage[T1]{fontenc}
\usepackage{hyperref}
\usepackage[margin=1in]{geometry}
\usepackage{amsmath,amssymb,amsthm}
\usepackage{graphicx}
\usepackage{booktabs}
\usepackage{multirow}
\usepackage{siunitx}
\usepackage{bbm}
\usepackage{diagbox}
\usepackage[ruled,vlined,linesnumbered]{algorithm2e}
\usepackage[table]{xcolor}
\usepackage[numbers,sort&compress]{natbib}
\usepackage{fontawesome5}
\usepackage{xurl}
\usepackage{enumitem}
\usepackage{subcaption}
\usepackage{tikz}
\usetikzlibrary{arrows.meta, positioning, fit, backgrounds, calc}

\usepackage{listings}
\usepackage{xcolor}
\usepackage{makecell}

\definecolor{codegreen}{rgb}{0,0.6,0}
\definecolor{codegray}{rgb}{0.5,0.5,0.5}
\definecolor{codepurple}{rgb}{0.58,0,0.82}
\definecolor{backcolour}{rgb}{0.95,0.95,0.92}

\lstdefinestyle{pythonstyle}{
backgroundcolor=\color{backcolour},
commentstyle=\color{codegreen},
keywordstyle=\color{magenta},
numberstyle=\tiny\color{codegray},
stringstyle=\color{codepurple},
basicstyle=\ttfamily\footnotesize,
breaklines=true,
numbers=left,
numbersep=5pt,
showstringspaces=false,
tabsize=4,
language=Python
}

\newtheorem{definition}{Definition}

\newcommand{\polynomials}{\ensuremath{a(x), b(x), a(y), b(y)}} % Will place : in the text \(a(x), b(x), a(y), b(y)\)
\newcommand{\mixedSixSix}{Base(66)}
\newcommand{\mixedSixSixSixNine}{Mixed 66-96}
\newcommand{\mixedSixSixSixNineOneTwoSix}{Mixed 66-96-126}
\newcommand{\mixedSixSixSixNineOneTwoSixOneFiveThree}{Mixed 66-96-126-153}
\newcommand{\groundAnchored}{Ground + Anchored(66)}%\_66-96-126-153}
\newcommand{\groundPure}{Ground}%groundPure\_66-96-126-153}
\newcommand{\bfit}[1]{{\fontseries{bx}\fontshape{it}\selectfont #1}}
\usepackage{pifont}

\definecolor{wongBlue}{RGB}{0,114,178}
\definecolor{wongGreen}{RGB}{0,158,115}
\definecolor{wongOrange}{RGB}{230,159,0}
\tikzset{
  archfont/.style={font=\sffamily\small},
  input/.style={draw=gray!60, rounded corners=2.5pt, fill=gray!6, align=center,
                inner sep=5pt, minimum height=10mm, text width=30mm},
  module/.style={draw=wongGreen, rounded corners=2.5pt, fill=wongGreen!10, align=center,
                 inner sep=5pt, minimum height=10mm, text width=28mm},
  pre/.style={draw=wongBlue, rounded corners=2.5pt, fill=wongBlue!8, align=center,
              inner sep=5pt, minimum height=10mm, text width=26mm},
  head/.style={draw=wongOrange!90!black, rounded corners=2.5pt, fill=wongOrange!14, align=center,
               inner sep=5pt, minimum height=10mm, text width=29mm},
  dims/.style={font=\sffamily\footnotesize, text=black!70, inner sep=1.5pt},
  tag/.style={font=\sffamily\footnotesize\scshape, text=black},
  arr/.style={->, semithick, draw=black!60},
  callout/.style={font=\sffamily\footnotesize, text=black!70, inner sep=1.5pt},
  lead/.style={draw=black!45, thin}
}

\begin{document}

\title{Designing Quantum Error Correcting Codes to fit decoders via Reinforcement Learning}

\author{Omer S. Sella}
\affiliation{Imperial College London}
\homepage{https://omer-sella.github.io/}
\email{o.sella@imperial.ac.uk}
\thanks{wishes to thank Dr. Panteleev Pavel for helpful clarifications, guidance and methodology in the early stages of this work.}
\orcid{0000-0002-2795-8580}
\author{Robert Pinsler}
\affiliation{Microsoft Research}
\orcid{0000-0003-1454-188X}
\author{Thomas Heinis}
\affiliation{Imperial College London}
\orcid{0000-0002-7470-2123}

\maketitle
\begin{abstract}
We present a reinforcement learning (RL) approach to the co-design of stabilizer sets of Quantum Error Correcting Codes (QECCs) and decoders. 
We show how to produce a generative model that produces Bivariate Bicycle (BB) codes based on the choice of decoder.
Specifically, we fix a decoder architecture and use Proximal Policy Optimisation (PPO) to train an agent over BB codes to maximise decoder performance under a depolarising channel noise model. 
\end{abstract}

\section{Introduction}\label{sec:intro}
Fault-tolerant quantum computation requires that physical noise rates be suppressed low enough, and decoding is done fast enough. 
Quantum error correcting codes (QECCs) achieve the former by restricting a larger space of physical 
qubits to a subspace of logical qubits that satisfy some relations.
These relations can be measured without collapsing the encoded quantum state. The design of such codes is just one side of a foundational problem in the path to scalable 
quantum computing. The choice of decoder is another, and the two operate together.
%For binary and Additive White Gaussian Noise (AWGN) channels, low-density parity check (LDPC) codes~\cite{gallager1962low} 
%were adopted in communication and data storage,  
%with a ranging array of decoding .
\noindent Quantum Low Density Parity Check codes (qLDPC)~\cite{panteleev2022asymptotically,bravyi2024high} present an attractive approach as they promise good asymptotic parameters and can be decoded using algorithms such as belief propagation (BP)~\cite{pearl1988probabilistic}.
Furthermore, the implication of \emph{low density} is that the connectivity required between logical qubits is low, an appealing property in some qubit architectures and quantum control hardware.
Practical decoding for fault tolerant quantum computing is likely to constrain the 
decoder design further than qubit connectivity.
Much like the binary setting, the decoder must complete its work under a 
certain time limit, otherwise its decoding latency may kill the entire application~\cite{sella2018fec}.

\noindent A central insight motivating this work is that code design and decoder design are not independent problems, 
especially in decoders that rely on heuristics and approximations of Maximum Likelihood (ML) decoding.
The performance of a given decoder depends strongly on the structure of the code.  
We can ask the question: \textit{given a fixed decoder, which code maximises its performance ?}\\

\noindent A deeper question that motivated this work is: \textit{Assuming deep knowledge of a quantum-device-under-test (qDUT), its error modalities, failure modes etc., how could one use that knowledge to engineer better, stack-integrated, QECC-decoder pair ?}

\noindent While we are still far from answering the latter, we believe this work presents some progress in answering the former. This paper proceeds as follows:\\
Section~\ref{sec:background} is meant to serve as a presentation of the tools we use in this work for a non expert in the corresponding field. We also include a \textbf{dataset} of BB codes we used for supervised learning in a later section.\\
Section~\ref{sec:mdp} describes a \textbf{Markov Decision Process} (MDP) formulation of Bicycle-Bivariate (BB) code design in which the state is a parity check matrix and an action modifies the matrix. Following the action we present a \textbf{reward} function based on the area under a fitted BER physical-error-rate curve, giving a single scalar.\\
\noindent A ``warm start'' strategy, in the form of a predictive model, is presented in Section~\ref{sec:critic}. It is then used in as part of the \textbf{policy architecture} in order to improve sampling efficiency.\\
\noindent Following the related work in Section~\ref{sec:related-work}, we evaluate the training and resulting policy in Section~\ref{sec:evaluation}, and discuss potential further work in Section~\ref{sec:conclusion}.

\section{Background\label{sec:background}}
In binary communication setting, a transmitter sends a codeword through a noisy channel, some of the bits may be corrupted, and the receiver attempts to reconstruct the message from the received bits. 
A binary linear code is defined\footnote{In a non-unique way.} by a parity check matrix \(H \in \mathbb{F}_2^{m \times n}\); codewords are elements of the null space \(\ker(H)\). LDPC codes have sparse \(H\), which enables efficient decoding via belief propagation (BP) on the Tanner graph~\cite{tanner1981recursive}, a bipartite graph in which bit nodes are connected to check nodes according to the nonzero entries of \(H\). BP then passes probabilistic messages along Tanner graph edges and terminates when all parity checks are satisfied or after a fixed number of iterations. 
Its performance degrades in the presence of short cycles in the Tanner graph. 
The soft information produced by BP algorithms can serve as a ranking of the indices in an increasing order of likelihood to be a non trivial Error. 
Using this ranking, a decoder can search for a \(rank(H)\) set of indices. If the corresponding rows of \(H\) happen to be linearly independent, the decoder can invert the sub-matrix and calculate the error.
Such sets of indices are called Information Sets~\cite{fossorier1995soft}, and the decoding method is termed Ordered Statistics Decoding (OSD). When the soft information used by OSD is produced by Belief-Propagation (BP), the decoding is often termed BP+OSD.

\noindent In the binary case, 
the error pattern is not the end-game, but rather a means to reconstruct the transmitted codeword
\footnote{Decoding would halt if at any point the sent codeword is recovered.}. 
The quantum setting is different, in the sense that we are not allowed to learn anything about the underlying quantum information, and the goal of the decoder is only to counter the error pattern, which is sometimes referred to as syndrome decoding. Keeping that in mind, we introduce qubits primarily as a backdrop for the errors that may experience.
\subsection{Qubits and Pauli operators\label{qubits}}
A qubit is a unit vector in \(\mathbb{C}^2\), which could be written as 
\begin{equation}
\begin{split}
|q\rangle = \alpha\binom{1}{0} + \beta\binom{0}{1} 
= \alpha\ket{0} + \beta\ket{1}
\end{split}
\end{equation}
where
\begin{equation}
\alpha,\beta \in \mathbb{C},\qquad |\alpha|^2 + |\beta|^2 = 1  
\end{equation}
A ``bit flip'' operator, \(X\), interchanges \(\ket{0}\) and \(\ket{1}\). A ``phase flip'' operator \(Z\) moves \(\ket{1}\) to \(-1\cdot \ket{1}\), and their composite, up to a global phase, is \(Y = iXZ\)\footnote{We remark that global phase, i.e., the \(i\) factor is disregarded.}:
\begin{equation}
X = \begin{pmatrix}0&1\\1&0\end{pmatrix}, \qquad
    Z = \begin{pmatrix}1&0\\0&-1\end{pmatrix}, \qquad 
    Y = \begin{pmatrix}0&-i\\i&0\end{pmatrix} \qquad
\end{equation}
If we assume that single-qubit errors are unitary linear maps, then %by choosing a basis for \({{\mathbb{C}}^2}^{\otimes n}\), meaning ; 
since \(X, Y, Z\) along with the identity matrix, \(I\) span the \(2\times 2\) complex matrices, any single-qubit noise channel can be written as a linear combination of the Pauli operators \(I,X,Z,Y\).
We also note that 
\begin{equation}
\begin{split}
    X\cdot Z = -1\cdot Z \cdot X\\
    X\cdot Y = -1\cdot Y \cdot X\\
    Z\cdot Y = -1\cdot Y \cdot Z\\
\end{split}
\end{equation}
and say that these operators anti-commute. A register (memory element) of \(n\) qubits resides in the tensor product of \(\mathbb{C}^2\), i.e., 
\({(\mathbb{C}^2)}^{\otimes n}\). For multi qubit errors the analysis is slightly more complicated, we explain next. 
It is worth knowing the formal definition of a Quantum Error Correcting Code~\cite{gottesman2026surviving}:
\begin{definition}[Quantum error-correcting code]
A quantum error-correcting code \(U,\mathcal{E}\) is a partial isometry
\(U:\mathcal{H}_{K}\to\mathcal{H}_{N}\)
with a set of correctable errors \(\mathcal{E}\), where each
\(E\in\mathcal{E}\) is a linear map \(E:\mathcal{H}_{N}\to\mathcal{H}_{M}\), such that there exists a recovery operation
\(\mathcal{D}:\mathcal{H}_{M}\to\mathcal{H}_{K}\) with
\(\forall E\in\mathcal{E},\ \forall|\psi\rangle\in\mathcal{H}_{K},\)
\begin{center}
\(\mathcal{D}(E U|\psi\rangle\langle\psi| U^{\dagger} E^{\dagger}) = c(E,|\psi\rangle)|\psi\rangle\langle\psi|.\)
\end{center}
\end{definition}
\noindent We note that this definition suggests we need to consider the noise from the very beginning. In this work, we consider the noise to be in the form of \(n\) dimensional Pauli operators, which can be described as tensor products of \(n\) (single qubit) Pauli operators (together with the identity matrix):
\begin{equation}
    P = P_0 \otimes\ldots\otimes P_{n-1} , \qquad P_i\in\{I,X,Z,Y\}
\end{equation}
For two operators \(M, N\) of the above form, their product is defined as the tensor product of their coordinate-wise product:
\begin{equation}
    E = \bigotimes_{i=0}^{n-1}{(M_i \cdot N_i)}
\end{equation}
We say that such operators commute if 
\begin{equation}
    E = \bigotimes_{i=0}^{n-1}{(M_i \cdot N_i)} = \bigotimes_{i=0}^{n-1}{(N_i \cdot M_i)}
\end{equation}
which, for Pauli operators, amounts to the number of coordinates where they anti-commute being even. For \(n\) dimensional Pauli operators, we can define two homomorphisms onto binary vectors:
\begin{definition}\label{eq:projection}
Let \(P\) be a dimension \(n\) Pauli operator \(P = P_0 \otimes \ldots \otimes P_{n-1} \) define two vectors, \(P_X\) and \(P_Z\)  with binary coordinates as follows:
\begin{equation}
\begin{split}
 P_X(k) = 0  \quad &if \quad P(k)= I\\
P_X(k) = 1   \quad &if \quad P(k)\in \{X,Y\} \\
 P_Z(k) = 0  \quad &if \quad P(k)= I\\
P_Z(k) = 1   \quad &if \quad P(k)\in \{Z,Y\} \\
\end{split}
\end{equation}
\end{definition}
\noindent The decomposition above respects multiplication, i.e., if \(P,Q\) are two \(n\) dimension Pauli operators, and \(P\cdot Q=S\) then:
\begin{equation}
    \begin{split}
        S_X =& P_X + Q_X \mod 2\\
        S_Z =& P_Z + Q_Z \mod 2
    \end{split}
\end{equation}
and therefore, to tell if \(P\) and \(Q\) commute is the same as checking that\footnote{This is very useful for leveraging binary error correcting codes (and decoders) and porting them to the quantum setting.}:
\begin{equation}
    \sum\limits_{k}P_X(k) \cdot Q_Z(k) +  \sum\limits_{k}P_Z(k) \cdot Q_X(k) = 0 \mod 2
\end{equation}
\noindent In the classical setting, the codewords can be seen as the null space of the parity check matrix and (recoverable) errors take a codeword to a non codeword, so out of this null space. The parity matrix, can, in turn be thought of as a stack of single parity checksums (SPCs). In the quantum setting, at least for some types of codes, we replace the stack of SPCs (rows of the parity check matrix) with a stack of Pauli operators, and we can look at the concurrent \(+1\) eigen space of these Pauli operators (which may be trivial). In the case this code is not trivial, this stack of Pauli operators are referred to as the \emph{stabilizers} of the code (since the stabilize, or fix, the space point-wise). It can be shown that if the code is non trivial, then all stabilizers commute. Calderbank-Steane-Shor (CSS) codes~\cite{calderbank1996good, steane1996error}, are such codes, in which every stabilizer contains only only \(X\) or \(Z\) components. This means that, for each operator in the stabilizer stack, only one of the projections~\ref{eq:projection} is non trivial, which allows us to look at two binary matrices, \(H_x, H_z\). This also provides an easy view into the code dimension, i.e.:
\begin{equation}\label{eq:codeDimension}
    N - rank(H_x) - rank(H_Z)
\end{equation}
The requirement that the stabilizers commute, is then translated to
\begin{equation}\label{eq:dualContaining}
    Hx \cdot (Hz)^T = 0
\end{equation}
Given any \(n\) dimensional Pauli operator \(E\), it either commutes or anti-commutes with each operator in the stack of stabilizers.
Suppose it has only \(X\) components, then it will commute with every \(X\) type stabilizer, or equivalently:
\begin{equation}
    P_X(E) \in ker(H_X)
\end{equation}
If it happens to have non trivial \(X\) components in an odd number of places as one of the \(Z\)-type stabilizers, then it anti-commutes with it, and so\footnote{and note that this is a binary vector}:
\begin{equation}
    s_Z = H_Z \cdot P_X(E) \neq 0 
\end{equation}
Decoding could then be thought of as the task of inferring 
\[ P_X(E) = E_X \text{ from }s_Z \]
and
\[ P_Z(E) = E_Z \text{ from } s_X\]
and reassembling an error $E'$ with the same syndrome. 
In fact, this powerful description of Pauli noise as well as the stabilizers as binary vectors,
readily yields a decoding algorithm. 
Starting from any binary BP algorithm that provides an error approximation, 
decode \(\hat{E}_Z\) from \(s_X\), \(\hat{E}_X\) from \(s_Z\), 
and return \(\hat{E}=P_X^{-1}(\hat{E}_X) \cdot P_Z^{-1}(\hat{E}_Z)\), i.e., the coordinate-wise multiplication of the inverse image of \(\hat{E}_X\) and \(\hat{E}_Z\)\footnote{An immediate heuristic from (potentially) a composite of heuristics.}.

\begin{figure}[htbp]
    \centering
    \includegraphics[width=0.5\linewidth]{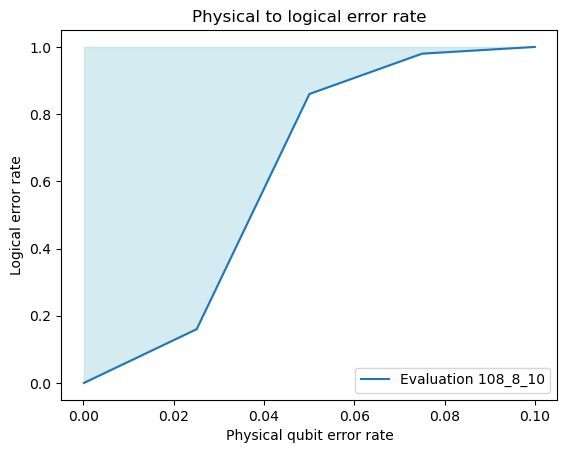}
    \caption{Physical-to-logical error rate curve explanation. The reward is calculated as the area between the performance curve and the constant function \(1\).}
    \label{fig:reward-explained}
\end{figure}

\noindent Given a code with parity matrices \(H\) and a decoder, a monte-carlo evaluation can be carried as in Algorithm~\ref{alg:monte-carlo}.\\
\noindent The reader with background in classical error correction will likely distinguish between \bfit{decoder failure}, where the decoder admits it \emph{failed} to decode, but is aware of this fact, and \emph{decoder error}, where the decoder thinks it found a correction, and did in fact not (making this a silent data corruption event). In our context, these events are counted equally.
A physical error rate to logical error rate can subsequently be drawn as in Figure~\ref{fig:reward-explained}.
In this work, we focus on training the agent at the physical error range 
\[0.001, 0.00316228, 0.01, 0.03162278, 0.1\]
corresponding to 
\begin{lstlisting}
numpy.geomspace(0.001, 0.1, 5)
\end{lstlisting}
the majority of the dataset introduced in section~\ref{sec:dataset}, however, was evaluated on a larger range (but a uniform number of samples across the error range). 

\subsection{Bivariate Bicycle Codes}
\label{sec:bb}
Bicycle codes are a sub-family of CSS codes, originally defined in~\cite{mackay2004sparse}. Bivariate Bicycle codes were subsequently defined as follows:
\begin{definition}[Bivariate Bicycle code~\cite{bravyi2024high}]
    \label{def:bbcode}
Let \(S_\ell\) be the cyclic shift matrix of size 
\(\ell\times\ell\) and set 
\begin{equation}
X = S_\ell \otimes I_m,\qquad Y = I_\ell \otimes S_m
\end{equation}
And note that \(XY=YX\) and \(X^l=Y^m=I_{lm}\). A BB code is specified by a pair of matrices
\begin{equation}
    A = A_1 + A_2 +A_3, \qquad B=B_1+B_2+B_3
\end{equation}
where each matrix \(A_i\) and \(B_j\) is a power of \(X\) or \(Y\), and subsequent parity check matrices, 
\begin{equation}
    H^X = [A \mid B], \qquad H^Z = [B^\top \mid A^\top].
\end{equation}
\end{definition}
\noindent The definition in~\cite{bravyi2024high} also suggested that the \(A_i\)s should be distinct as to not cancel out (similarly for the \(B_i\)s). 
In this work, we take a slight generalisation of this definition, following~\cite{voss2024multivariate, voss2025multivariate}, by defining polynomials \polynomials{} and subsequently taking:
\begin{equation}
A=a(X) \oplus a(Y)  \qquad B = b(X) \oplus b(Y)     
\end{equation}
\noindent Either definition guarantees the commutation of the resulting stabilizers, since circulant matrices (\(X,Y\) and their powers) commute, as well as a symmetry between the number of \(X\) and \(Z\) stabilizers. The code has length \(n = 2l m\). The code is entirely determined by the coefficients of the four polynomials \polynomials{}. 
This compact parameterisation is important to our RL formulation: 
rather than searching over binary matrices, each of size \((lm) \times (2lm)\), 
the agent searches over a space of four polynomials of degrees at most \(l-1,m-1\) and binary coefficients. 
The reader should be aware that other definitions, narrower and wider, exist in~\cite{panteleev2021degenerate, wang2024coprime, eberhardt2024logical, kovalev2013quantum}.

\subsection{Proximal Policy Optimisation}
\label{sec:ppo}
Loosely speaking, reinforcement learning provides a framework for problems where we make sequential decisions, or ``actions'', 
moving us through ``states'', each associated with a ``reward''.
If we denote the state at time \(t\) by \(s_{t}\) and the action by \(a_{t}\), then a trajectory \(\tau\), 
is a sequence of states and actions, starting at \(t=0\), i.e.:
\begin{equation}
\tau = \{s_i, a_i\}_{i=0}^{\infty}
\end{equation}

\noindent If we further denote the reward by \(r_t(s_{t}, a_{t})\)
then we can look at the discounted cumulative reward:
\begin{equation}
\label{eq:return}
 R(\tau) = \sum_{i=0}^{\infty} \gamma^{k}\cdot r_t 
\end{equation}

\noindent Assuming one starts from a state \(s\), and acting using a policy \(\pi\), we can look at the following two functions: the first is the value function, which tells what is the value of acting using \(\pi\) from state \(s\):
\begin{equation}
    V^{\pi}(s) = \underset{\tau \sim \pi}{E}\left[R(\tau)| s_0 = s\right]
\end{equation}
The second is: if we were to take an action \(a\) (not necessarily using the policy \(\pi\)), what would be the value of continuing the trajectory but sticking to the policy \(\pi\), i.e., the Action-Value function:
\begin{equation}
    Q^{\pi}(s,a) = \underset{\tau \sim \pi}{E} \left[R(\tau)| s_0 = s, a_0 = a\right]
\end{equation}
\noindent The advantage function, is then defined by:
\begin{equation}
A^{\pi}(s,a) = Q^{\pi}(s,a) - V^{\pi}(s)
\end{equation}

\noindent The goal of an RL agent is to learn a parametrized policy, \(\pi_{\theta}\) that maximises the return~\ref{eq:return}.
Policy gradient methods are a family of RL algorithms that learn such policies, and are appealing especially 
when the action and state spaces are either continuous or high-dimensional discrete.
In this work we use Proximal Policy Optimisation (PPO)~\cite{schulman2017proximal}, which forces a policy 
update to stay close to the current policy via a clipped surrogate objective:
\begin{equation}
\begin{split}
    &L^{\text{CLIP}}(\theta) = \\
    &\mathbb{E}_t\!\left[\min\!\left(\rho_t(\theta)\hat{A}_t,\; \text{clip}(\rho_t(\theta), 1-\epsilon, 1+\epsilon)\hat{A}_t\right)\right]
\end{split}
\end{equation}
where \(\rho_t(\theta) = \pi_\theta(a_t|s_t)/\pi_{\theta_{\text{old}}}(a_t|s_t)\) is the importance weight and \(\hat{A}_t\) is an approximation of the advantage function.
In addition to the policy \(\pi_\theta\), PPO also learns an approximation to the value 
function \(V_\phi(s)\). The overall algorithm is summarised in Algorithm~\ref{alg:main}. 
The code we used to apply PPO, once we state the problem in terms of states and actions in Section~\ref{sec:mdp}, 
is almost a drop-in-place reuse of the code explained in~\cite{bou2024torchrl}, and described by Algorithm~\ref{alg:main}.

\subsection{Dataset\label{sec:dataset}}
We accumulated a dataset of BB codes of various parameters as described in Table~\ref{tab:recordCensus}. 
Although it is not strictly needed to carry out PPO training in Algorithm~\ref{alg:main}, 
we used it in supervised learning manner, in an effort to start PPO from a better-than-random policy weights. 
We discuss this strategy, as well as our findings in Section~\ref{sec:critic}.
The collection was done by setting the agent to high exploration mode (by enforcing high entropy) and setting the environment to log them.
\begin{figure}[htbp]
    \centering
    \includegraphics[width=0.9\linewidth]{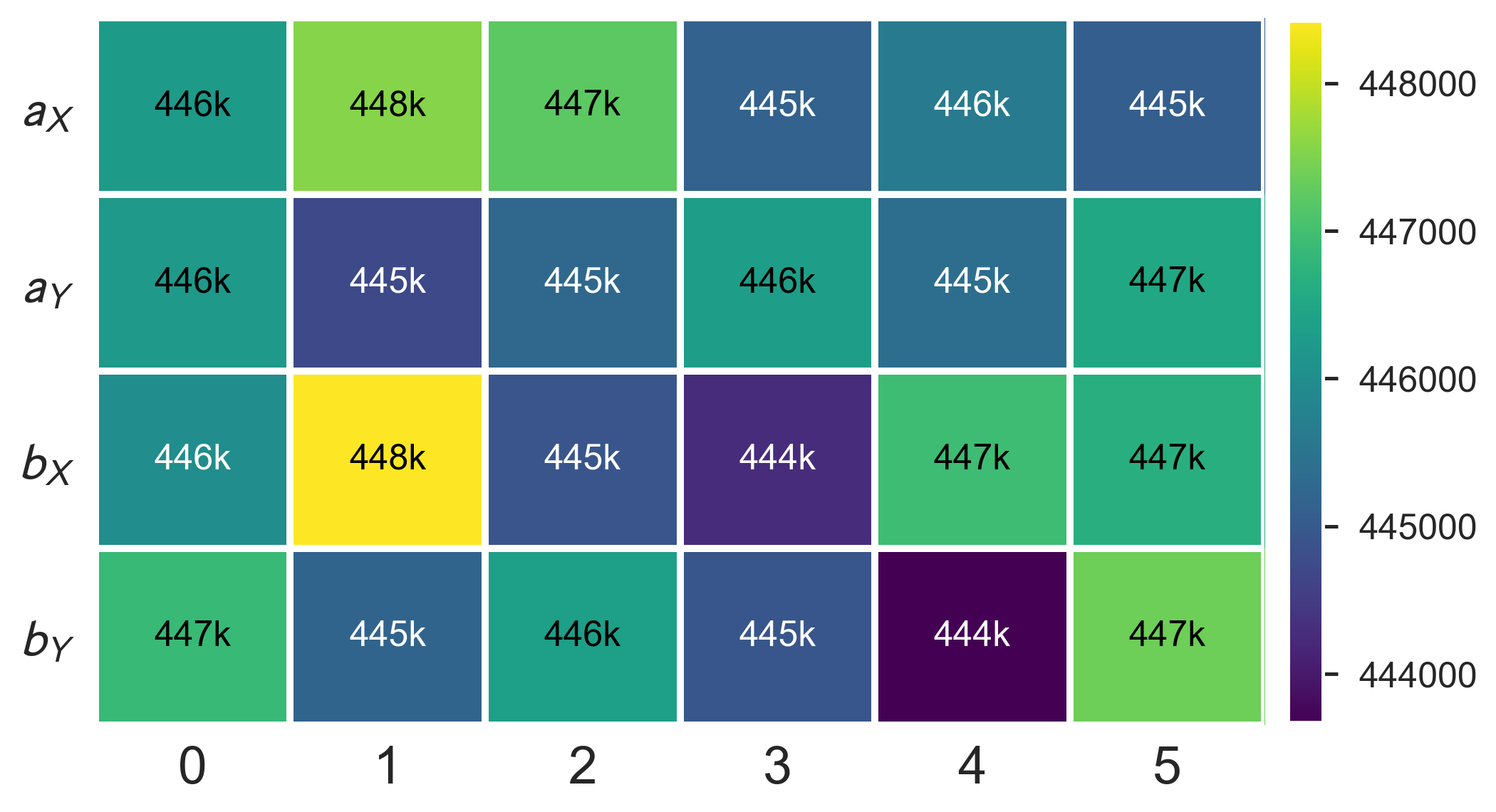}
     \label{fig:coefficientsHeatmap-6-6}
     \caption{Coefficient heatmap showing hits per coefficient index for the codes in the \(l=6, m=6\) dataset. Similar figures for other code sizes are provided in~\ref{appendix:dataset}, Figure~\ref{fig:coefficientsHeatmap-rest}.}
\end{figure}
A more extensive statistical analysis of this dataset can be found in Appendix~\ref{appendix:dataset}. 
The dataset itself, as well as the code used to produce the figures describing is available for public use~\cite{sella2026bbcodesdataset}.
The dataset provides ``labels'', i.e., the monte-carlo evaluation of each code over \(50\) samples per point. 
A typical .jsonl file in the repository contains multiple code records, each with the code parameters 
(\(l,m\), \polynomials{}, number of logical qubits) 
as well as the error range over which the code was evaluated, and the corresponding logical error counts. 
We note that the codes were not evaluated until a fixed number\footnote{The usual methodology is to observe \(100\) errors for high confidence.} 
of errors were observed, but rather over a fixed number of samples, as we use in training.

\begin{figure}[htbp]
    \centering    
    \includegraphics[width=0.8\linewidth]{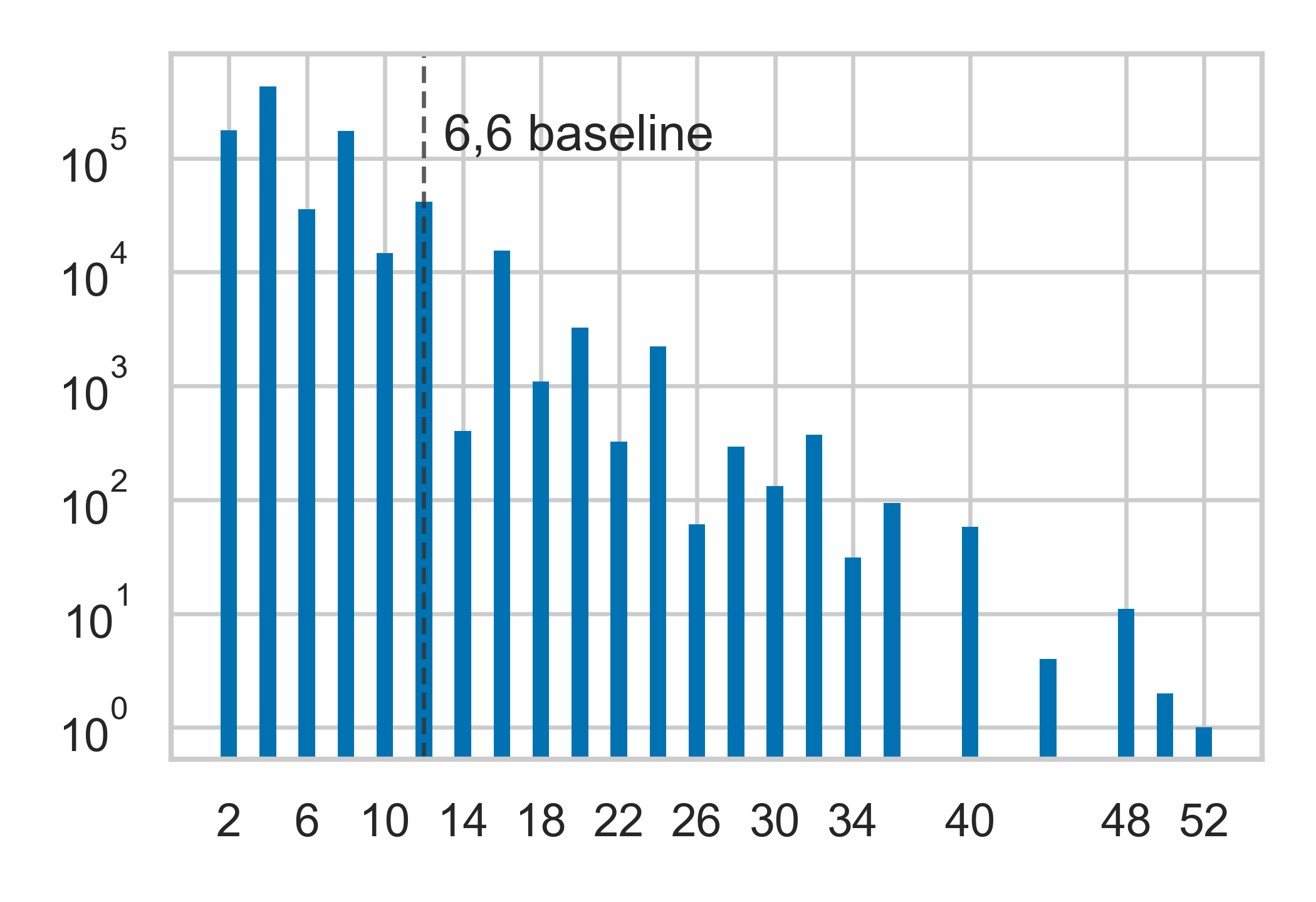}
  \caption{Distribution of the number of logical qubits for the codes in the \(l=6, m=6\) dataset. Similar figures for other code sizes are provided in~\ref{appendix:dataset}, Figure~\ref{fig:kDistribution-rest}}
  \label{fig:kDistribution-6-6}
\end{figure}

\noindent It should be stated that, the representation of a BB code using the polynomials \polynomials{} is not unique, 
and the dataset contains duplicates. Two evaluations of the same code by the same decoder are independent Bernoulli trials of the same probability, so pooling them could improve the estimate.

\section{Framing as a sequential decision making problem}\label{sec:mdp}
For a code, defined by a pair of parity matrices \(H_X, H_Z\) and a choice of decoder, we define performance based on the evaluation~\ref{alg:monte-carlo} and the resulting plot~\ref{fig:reward-explained}.
The first problem with the way the evaluation data is gathered is that it is only a noisy approximation of the true performance of the code. 
This can be accommodated in learning paradigms like PPO.
The second, has to do with the way it is presented in Figure~\ref{fig:reward-explained}, namely, 
that it contains multiple scalar values. 
A ``better'' code will have a steeper fall, and its curve stays lower. We collapse it into a single scalar in the following section.
\begin{algorithm*}
\DontPrintSemicolon
\caption{A Monte Carlo evaluation of code-decoder pair}
\label{alg:monte-carlo}
\KwIn{\(H_X,\quad H_Z,\quad errorRange,\quad numberOfSamples\)}
\KwOut{\(logicalErrorRate\)}
Prepare \(X\) and \(Z\) logical operators \(L_X,\quad L_Z\) from \(Hx, Hz\)\;
\For{ \(p\) in errorRange }{
     \For {i in  1..numberOfSamples } 
     { 
         Sample \(E\) with probability \(p\) for \(E(i)\in\{X,Z,Y\}\) \;
         Calculate \(s_X = P_Z(E)\cdot H_X,\quad s_Z = P_X(E)\cdot H_Z\)\;
         \(\hat{E_X},\hat{E_Z}\) = decode\((H_X,H_Z,s_X,s_Z)\) \;      
         \(R_X \gets (\hat{E}_X + E_X) \bmod 2\), \quad \(R_Z \gets (\hat{E}_Z + E_Z) \bmod 2\)\tcp*{residual error} \;
         \If{\(H_X \cdot R_Z \neq 0\) or \(H_Z \cdot R_X \neq 0\)}
         {logicalErrorRate \(+= 1\)}\tcp{The residual error does not commute with the stabilizers, i.e., nonzero syndrome}
         \Else{
         \If{\(L_X \cdot R_Z \neq 0\) or \(L_Z \cdot R_X \neq 0\)}
         {logicalErrorRate \(+= 1\)} \tcp{The residual error does not commute with the logical operators, i.e., it is not a stabilizer}}
         }}
\end{algorithm*}

\subsection{Reward}
%We can encompass this (albeit not perfectly) using integration.
To obtain a single scalar that is function of this scattered dataset, integrate to obtain the area bounded between the evaluation curve and the constant function \(1\)\footnote{We note that this is a design choice, and one could choose differently (for example, by sampling a single error point).}.
\begin{equation}
\label{eq:reward}
    r(H_X,H_Z)=\int_{p_{\min}}^{p_{\max}}\big(1-\widehat{f}(p;H)\big)dp
\end{equation}
where \(\widehat f\) is the empirical combined failure rate, computed by the trapezoid rule over the error range and 
where \([p_{\min}, p_{\max}]\) is the evaluation range.
% Larger (less negative) reward corresponds to better code performance.
This reward has several desirable properties, and some that are not desirable.
While a higher number would mean ``less errors observed in the range'', the integration provides a very small number, which we compensate for by dividing it by the width of the error range.
Another problem with the above reward, as it stands, is that it rewards the agent solely on logical error rate. 
A code with no logical qubits will be highly performant by this definition, 
and this could drive the agent towards codes with fewer logical qubits, denoted as \(k\).
One way to discourage any agent negatively on coming up with a code that has no, or not enough, logical qubits, is to engineer the reward function to 
be proportional to the distance of \(k\) from some minimum requirement on the number of logical qubits we expect a code should have, denoted \(k_{min}\), i.e.:
\[(k-k_{min})/k_{min}\]
For \[k \leq k_{min}\]
and switch to the decoder generated evaluation when \(k \geq k_{min}\). We summarise this as Definition~\ref{def:reward}:
\begin{definition}[Reward\label{def:reward}]
  \begin{equation}
r(H_X,H_Z) =
       \begin{cases}
                    \frac{1}{p_{\max} - p_{\min}}\int_{p_{\min}}^{p_{\max}}\big(1-\widehat{f}(p;H)\big)dp & \text{if } k \geq k_{min},\\[1ex]
                     \frac{k-k_{min}}{k_{min}}& \text{if } k < k_{min}.
        \end{cases}
\end{equation}
\end{definition}
\noindent Before moving ahead, we note that in many constructions the number of logical qubits can be calculated, and even designed by the choice of the polynomials. 
We chose to treat it as a quantity that is to be learned by the agent, since other constructions may not allow this quantity to be designed.

\subsection{Framing as a Markov Decision Process}
At first glance, we could state the problem as ``stateless'', i.e.,  
for given \(l,m\), produce polynomials \polynomials{}, and construct code matrices from them as in~\ref{def:bbcode}.
This has no consideration for the ``present state code'', whereas in the long run, 
we would like a generative model that would augment an existing code as little as possible, to yield a better performing code, maybe even just one term of one polynomial at a time.
%We formulate the problem as a Markov Decision Process \((\mathcal{S}, \mathcal{A}, r, \mathcal{T})\).
To this end, 
\paragraph{State} The state \(s \in \mathcal{S}\) are the current parity check matrices  
\(H^X = [A \mid B], \qquad H^Z = [B^\top \mid A^\top]\), where \(A,B\) are generated from the polynomials \polynomials{}.

\paragraph{Transition} The transition is deterministic:
\begin{equation}
    \begin{split}
    H' =& \mathcal{T}(H(\polynomials{}),\\
        &A^{a(x)},A^{b(x)},A^{a(y)},A^{b(y)})
    \end{split}
\end{equation}
 applies the action and returns the modified matrix.

\paragraph{Action} An action \(A^p\) on the polynomial \(p(x) = c_0 + c_1 \cdot x + \ldots\), where \(c_i \in \{0,1\}\), selects at most one index \(i\) and XORs the coefficient of \(c_i\) with \(1\), i.e., \(c_i \oplus 1\).
We subsequently define an action on the polynomials \polynomials{} to be a tuple of individual actions \(A^{a(x)},A^{b(x)},A^{a(y)},A^{b(y)}\). We will abuse notation and write 
\begin{equation}
    p(t)\oplus A^{p(t)}
\end{equation}
instead of explicitly saying which coefficient of \(p(t)\) is being flipped.

\paragraph{Reward} After each action, the new matrices \({H^{X}}', {H^{Z}}'\) are constructed, their rank is calculated, and 
the reward \(r({H^{X}}', {H^{Z}}')\) from Definition~\ref{def:reward} is returned.

\paragraph{Episode structure} Each episode begins from a random set of \polynomials{}, with up to \(3\) non zero coefficients, and proceeds for a fixed number of transitions. There are subtle failure modes which sit on the boundary of the hyperparameters (specifically the entropy coefficient), error range, code parameters and reward. First, we note that an actor which applies only one action (which could happen for example, if policy entropy collapses) is either stuck on a single code, or toggles between two codes. The single code case is exactly when the single action is ``do nothing'', communicated through a per-polynomial ``no-op'' bit to the environment. Otherwise there is at least one \(p(t)\in \{\polynomials\}\) with non trivial \(A^{p(t)}\). Say, for example, \(p(t) = b(x)\). Following action \(A^{b(x)}\) the new polynomial is \(b(x)\oplus A^{b(x)}\). Following another identical action \(b(x)\oplus A^{b(x)}\oplus A^{b(x)}\) is again \(b(x)\). 
This becomes catastrophic for a certain combination of BB code parameters and agent-environment hyperparameters. In error ranges that are too high, almost no code succeeds. In physical error rates that are too low, a fixed number of samples in Algorithm~\ref{alg:monte-carlo} provides a limited view, since the error sampled may be trivial, and this changes as a function of the code length. This may lead to mediocre codes that score a positive, yet lower-than-possible reward\footnote{Think of a code where the number of logical qubits is the same as the number of physical qubits, under low enough error rate.}. The agent could then get stuck on such mediocre codes, at the expense of exploration. To counter that effect, we cap each episode at \(\mathcal{O}(l+m)\) steps, at the end of which the environment communicates a ``truncated'' signal\footnote{which is different than a ``terminated'' signal~\cite{towers2024gymnasium}.} This means that, over the span of collection, the agent will have experienced a sufficiently diverse view~\cite{sutton2018reinforcement}.

\begin{algorithm*}
\DontPrintSemicolon
\caption{RL-based QECC co-design }%\href{https://github.com/Omer-Sella/qecc-rl/blob/main/src/qecc/reinforcementLearning.py}{\faGithub}}
\label{alg:main}
initialise policy \(\pi_\theta\) and value network \(V_\phi\)\;
\For{\(\mathrm{batch} = 1, 2, \ldots\)}{
  \For{each of the \(F\) transitions in the batch}{
    \tcp{episodes reset each polynomial to a random vector with at most
         three non-zero coefficients, and truncate after \(T\) steps}
    sample \(A^{aX}\!, A^{bX}\!, A^{aY}\!, A^{bY} \sim \pi_\theta(\cdot \mid s)\), each one-hot (includes no-op)\;
    \(p \gets p \oplus A^{p}\) for each polynomial %\(p\)\tcp*{\href{https://github.com/Omer-Sella/qecc-rl/blob/739f5cd/src/qecc/bb_gym_v_0_1.py\#L195}{\texttt{step()}}}
    build \(H'\); \(k \gets\) number of logical qubits\;
    \eIf{\(k \ge k_{\min}\)}{
      \(r \gets\) Monte-Carlo decoder evaluation of \(H'\)\;
    }{
      \(r \gets (k - k_{\min})/k_{\min}\)\tcp*{no decoding performed}
    }
  }
  update \(\pi_\theta, V_\phi\) with PPO on the batch\tcp*{truncated episodes bootstrapped}
  record the best \(H\) found so far\;
}
\KwRet{policy and value weights, evaluation log}
\end{algorithm*}

\section{Policy Architecture}
\label{sec:policy}
The only ingredient missing in order to apply Algorithm~\ref{alg:main} is a parameterisation of the
policy \(\pi_\theta\) and of the value function \(V_\phi\).
Both read the same state, i.e., the pair \(H^X = [A \mid B],\ H^Z = [B^\top \mid A^\top]\) built from the
polynomials \polynomials{}, 
but are required to answer different questions. The actor is asked which
action to take next, which in our framing corresponds to ``which coefficient of each polynomial to flip next'' (if any). 
The critic is meant to estimate what is the value of the present code, with the present policy.
It helps to think of the case where we have arrived at ``the best possible'' performing code. In such a
case we would like the actor to choose to do nothing, and the critic to estimate a value that is directly related 
to how well that code performs under the decoder. 
The critic therefore has to approximate, implicitly, what Algorithm~\ref{alg:monte-carlo} approximates explicitly, which is to map a code to its performance.

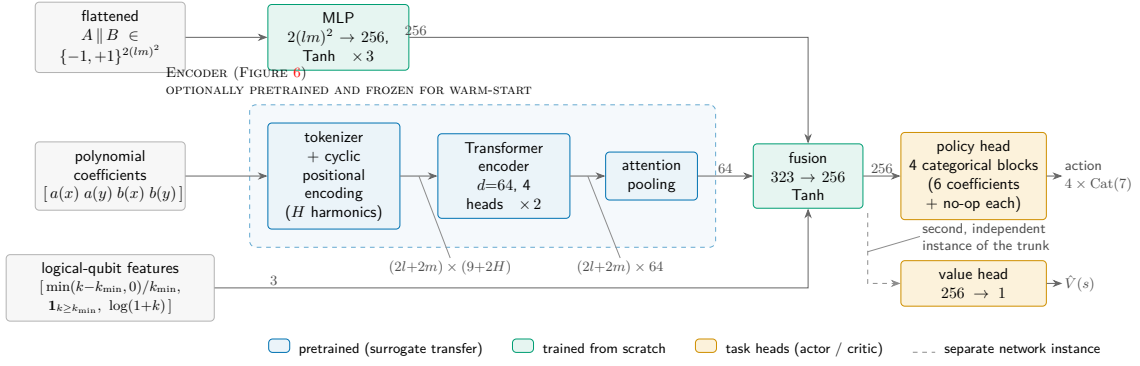
\begin{figure*}
  \centering
  \resizebox{\textwidth}{!}{\begin{tikzpicture}[archfont, >={Stealth[length=2.2mm]}]

\node[input] (flat) at (0, 0)     {flattened\\ $A\,\Vert\,B \in \{-1,+1\}^{2(lm)^2}$};
\node[input] (poly) at (0, -3.1)  {polynomial coefficients\\ $[\,a(x)\; a(y)\; b(x)\; b(y)\,]$};
\node[input, text width=43mm, minimum height=13mm] (kfeat) at (0, -5.6)
     {logical-qubit features\\
      {\footnotesize $[\,\min(k{-}k_{\min},0)/k_{\min},$\\
       $\mathbf{1}_{k\ge k_{\min}},\ \log(1{+}k)\,]$}};

\node[module] (mlp) at (5.1, 0) {MLP\\ \(2(lm)^2 \to 256\), Tanh\quad \(\times\,3\)};

\node[pre] (tok)  at (5.0, -3.1)  {tokenizer + cyclic\\ positional encoding\\ ($H$ harmonics)};
\node[pre] (trf)  at (8.8, -3.1)  {Transformer encoder\\ $d{=}64$, 4 heads\quad $\times\,2$};
\node[pre, text width=17mm] (pool) at (12.1, -3.1) {attention\\ pooling};

\begin{scope}[on background layer]
\node[draw=wongBlue!60, dashed, rounded corners=4pt, fill=wongBlue!4,
      fit=(tok)(trf)(pool), inner sep=4mm] (encbox) {};
\end{scope}
\node[tag, align=left, anchor=south west] at ($(encbox.north west)+(-20mm,1.0mm)$)
   {Encoder (Figure~\ref{fig:encoderArchitecture}) \\
   optionally pretrained and frozen for warm-start};

\node[module, text width=21mm, minimum height=14mm] (fuse) at (15.6, -3.1)
     {fusion\\ \(323 \to 256\)\\ Tanh};

\node[head] (actor)  at (19.3, -3.1) {policy head\\ 4 categorical blocks\\ (6 coefficients + no-op each)};
\node[head] (critic) at (19.3, -5.5) {value head\\ $256 \to 1$};
\node[dims, right=3.5mm of actor, align=left] (act) {action\\ $4 \times \mathrm{Cat}(7)$};
\node[dims, right=3.5mm of critic] (val) {$\hat V(s)$};

\draw[arr] (flat) -- (mlp);
\draw[arr] (mlp.east) -- node[dims, above, pos=0.25]{$256$} ++(6mm,0) -| (fuse.north);
\draw[arr] (poly) -- (tok);
\draw[arr] (tok) -- coordinate[midway] (m1) (trf);
\draw[arr] (trf) -- coordinate[midway] (m2) (pool);
\draw[arr] (pool.east) -- node[dims, above]{$64$} (fuse.west);
\draw[arr] (kfeat.east) -- node[dims, above, pos=0.1]{$3$} (kfeat.east -| fuse.south)
           -- (fuse.south);

\draw[arr] (fuse.east) -- node[dims, above]{$256$} (actor.west);
\draw[arr] (actor) -- (act);

\coordinate (cstart) at ($(fuse.south east)+(1mm,-1mm)$);
\draw[arr, dashed, draw=black!45] (cstart) |- (critic.west);
% placed in the clear gap between the two heads: the old position at (16.7,-4.5)
% was straddled by the k-feature arrow and by the dashed critic branch
\node[callout, align=left, anchor=west] (inst) at (18.1, -4.45)
     {second, independent\\ instance of the trunk};
\draw[lead] (inst.west) -- ($(cstart)!0.55!(cstart |- critic.west)$);
\draw[arr] (critic) -- (val);

\node[callout] (c1) at (7.6, -5.15)  {$(2l{+}2m)\times(9{+}2H)$}; %this is the callout for the first encoder bus (datapath)
\draw[lead] (c1.north) -- (m1);

\node[callout] (c2) at (11.4, -5.15) {$(2l{+}2m)\times 64$}; %this is the callout for the second encoder bus (datapath)
\draw[lead] (c2.north) -- (m2);

\node[anchor=north west, align=left, font=\sffamily\footnotesize, text=black]
  at (3.4, -6.6)
  {\tikz{\node[pre, text width=4mm, minimum height=3mm, inner sep=1pt]{};}\; pretrained (surrogate transfer)\qquad
   \tikz{\node[module, text width=4mm, minimum height=3mm, inner sep=1pt]{};}\; trained from scratch\qquad
   \tikz{\node[head, text width=4mm, minimum height=3mm, inner sep=1pt]{};}\; task heads (actor / critic)\qquad
   \tikz{\draw[dashed, black!45, semithick] (0,0) -- (5mm,0);}\; separate network instance};

\end{tikzpicture}}
  \caption{Network architecture. Each of the actor and critic instantiates
  the architecture independently. The structure encoder
  (blue) has the ability to be warm-started as explained in Section~\ref{sec:critic} and kept frozen for the
  first \(N\) collector batches while the randomly initialised branches
  (teal/orange) settle. 
  The fusion input size \(323 = 64_{\mathrm{enc}} + 256_{\mathrm{mlp}} + 3_{k}\) is
independent of the code size. The MLP input width \(2(lm)^2\), the token count
\(2l+2m\), and the policy-head sizes \(l{+}1,\ m{+}1\) scale with \(l,m\). All
dimensions in the figure are shown for \(l=m=6\) and \(H=3\).}
  \label{fig:hybrid}
\end{figure*}

\noindent Because the environment constructs the code from the polynomials, the same state is available to us in
more than one form, and at no extra cost:
\begin{enumerate}
  \item The coefficients of \polynomials{}, which are the compact description of the code, and how the action was defined.
  \item The expanded matrices \(H_X, H_Z\), which is what the decoder actually uses, and
  \item The number of logical qubits \(k\), which is easily computable from \(H_X,H_Z\) (by using Equation~\ref{eq:codeDimension}).
\end{enumerate}
In our policy architecture, each one its own branch, then ``fuse'' them using a single layer, 
on top of which there is either a single value head used by the critic, or parametrised Categorical distribution over the action space, used by the actor.
The resulting network is drawn in Figure~\ref{fig:hybrid}. 
The actor and the critic each instantiate it independently, i.e., they share the design but not the weights.
We first describe branch that consumes the code matrices, as it is straightforward.
%Section~\ref{sec:critic}, do we ask which parts of this network can be trained \emph{before} PPO begins.

\subsection{Multi-Layer-Perceptron (MLP)\label{sec:mlpBranch}}
The most direct parameterisation uses the matrices \(H_X,H_Z\) as the decoder uses them. 
Since these matrices are linear functions of \(A,B\) from Definition~\ref{def:bbcode}, we flatten the
concatenation of \(A,B\) and feed the resulting \(2{(lm)}^{2}\)-dimensional vector (\(2592\) entries for \(l=m=6\)) 
into a three-layer MLP with \(\tanh\) activations, which returns a \(256\)-dimensional vector.

\noindent The advantage of this part of the architecture is that it gives the agent complete visibility of the observed state, so it
does not need to learn a latent representation of what it is acting on. 
The disadvantage is double, as follows:
\begin{enumerate}
  \item It increases the size of the network, as a function of \(l,m\), and consequently the number of agent-environment
  interactions needed to learn. This is a costly sampling task even for codes of small size, since
  every interaction that produces a code with \(k \geq k_{\min}\) costs a full Monte-Carlo evaluation.
  \item The learning becomes code-size specific. The width of the input layer is \(2{(lm)}^{2}\), so the
  weights of a network trained at \((6,6)\) cannot be reused at \((9,6)\), and nothing learned at one
  size transfers to another.
\end{enumerate}
For these reasons we add a second branch, which is size transferable.

\subsection{BB code construction aware encoder\label{sec:encoderBranch}}
The second part of the architecture learns from the polynomials \polynomials{}, 
through an encoder\footnote{To clarify, the \bfit{encoder} in this context is a function that accepts code
parameters, and encodes them into a latent vector representation, to be used downstream (and not an
encoder of physical qubits to logical qubits).}, i.e., a fixed size neural network whose job is to turn the
parameters of a code (four polynomials as well as \(l,m\)) into a fixed-size latent vector that
is meant to predict the expected reward and code structure.
We design this encoder, and its learned representation, to be transferable across code sizes.
Later in this section we discuss under what circumstances this learned representation transfers, and how successfully.\\
      
 \noindent From a code's polynomials we produce tokens as described below, project each token
linearly to \(d=64\), pass the sequence through two transformer encoder layers
(\(d=64\), \(4\) attention heads, feed-forward width \(128\), GELU activations), and
pool the token representations by attention into a single \(64\)-dimensional latent
vector. 
All encoders in Section~\ref{sec:critic} share this architecture.
The number of harmonics \(H \in \{3,4,6,7,10\}\) is the only architectural variable.
    
\noindent While the number of tokens depends on the code size, the width of each token does not change, i.e., the same network can accept the tokens from any
code size as input.
This means the encoder could be regarded as a standalone objective that could be pretrained.
We review our approach and results in Section~\ref{sec:encoderEvaluation}. 
While we think of it as an integral part of this work,
the reader may choose to treat Section~\ref{sec:critic} as a side-quest, one that could be returned to later.
%We continue with \textbf{feature extraction}, which is where we rely on understanding the
%structure of BB codes as we defined them in~\ref{def:bbcode}.
  
  \noindent We have already capitalised on the compact representation of a BB code using the polynomials \polynomials{} to define the action space in~\ref{sec:mdp}.
  We now capitalise further on their cyclic property. Since~\ref{def:bbcode} implies \(x^{l}=y^{m}=I\), 
  each non zero binary coefficient naturally maps to a position on the unit circle using the transform:
  \begin{equation}
  %\begin{split}
   c_{k}\cdot x^{k} \rightarrow e^{k\cdot 2\cdot i \cdot \pi /p}
   % &=\cos(k\cdot 2\cdot i\cdot \pi/p) + i\sin(k\cdot 2\cdot i\cdot \pi/p)
  %\end{split}
  \end{equation}    
    
\noindent where \(i\) is a square root of \(-1\), and \(p \in \{l,m\}\).
  Exponentiating further by \(h\in \{1,2,3\ldots H\}\) we may obtain further expressiveness for a neural network
  to use.
  Each of the coefficients of the four polynomials \polynomials{} becomes one
  \textbf{token}.
  Table~\ref{tab:token-features} lists its features. Coefficient \(k\) of a
  polynomial with period \(p\) (\(p=l\) for \(a(x),b(x)\); \(p=m\) for \(a(y),b(y)\)) sits at
  angle \(\theta_k = 2\pi k/p\) on the unit circle, so the cyclic features respect the
  modulus \(x^{l}=y^{m}=I\).
  The \(H\) harmonics form a Fourier basis, and the first linear layer of the encoder learns its coefficients.
  So for a fixed choice of \(H\), each token has width \(F = 9 + 2H\).
  This makes every feature a \textbf{function} of \((k,p,l,m)\) that is a real number, 
  and the token \textbf{width} is independent of \(l,m\).
  
\begin{table*}[t]
  \centering
  \caption{Anatomy of one encoder token (width \(F = 9 + 2H\). At \(H=3\) \(F=15\)). A code
  of size \(l,m\) produces \(2l+2m\) such tokens, one per coefficient in the order
  \([a(x)\,|\,a(y)\,|\,b(x)\,|\,b(y)]\).}
  \label{tab:token-features}
  \footnotesize
  \begin{tabular}{llcl}
    \toprule
    Feature & Value & Width & Role \\
    \midrule
    Coefficient bit & \(\{0,1\}\) & 1 & is this monomial set to \(1\) \\
    One-hot label & \(e_g \in \{0,1\}^4\) & 4 & which polynomial from a(x), a(y), b(x), b(y)\\
    Cyclic position & \((\sin h\theta_k, \cos h\theta_k)\) & \(2H\) & position on the period-\(p\) cycle, per h=1\ldots H \\
    Linear position & \(k/p \in [0,1)\) & 1 & absolute position\\ %Omer: Note that this is not an unintended unbalance parenthesis problem !!!
    Global size features & \(\log(l),\; \log(m),\; 1.0\) & 3 & code size, identical across all tokens \\
    \bottomrule
  \end{tabular}
\end{table*}
  
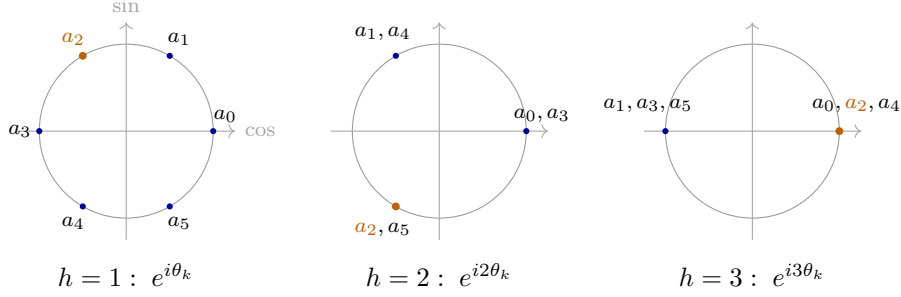
\begin{figure*}[!htbp]
  \centering
  \begin{tikzpicture}[scale=1.15]
    
    \begin{scope}[shift={(0,0)}]
      \draw[black!40] (0,0) circle (1);
      \draw[->,black!35] (-1.25,0)--(1.25,0) node[right,font=\scriptsize]{\(\cos\)};
      \draw[->,black!35] (0,-1.25)--(0,1.25) node[above,font=\scriptsize]{\(\sin\)};
      \foreach \k in {1,3,4,5}{
        \pgfmathsetmacro\ang{60*\k}
      \fill[blue!60!black] (\ang:1) circle (0.035);
      \node[font=\scriptsize] at (\ang:1.22) {\(a_{\k}\)};
    }
    \fill[blue!60!black] (0:1) circle (0.035);
    \node[font=\scriptsize] at (1.14,0.18) {\(a_{0}\)};
    \fill[orange!75!black] (120:1) circle (0.045);
    \node[font=\scriptsize,orange!75!black] at (120:1.22) {\(a_{2}\)};
    \node[font=\small] at (0,-1.65) {\(h=1:\; e^{i\theta_k}\)};
  \end{scope}
  
  \begin{scope}[shift={(3.6,0)}]
    \draw[black!40] (0,0) circle (1);
    \draw[->,black!35] (-1.25,0)--(1.25,0);
    \draw[->,black!35] (0,-1.25)--(0,1.25);
    \fill[blue!60!black] (0:1)   circle (0.035);
    \node[font=\scriptsize] at (1.18,0.18) {\(a_0,a_3\)};
    \fill[blue!60!black] (120:1) circle (0.035);
    \node[font=\scriptsize] at (120:1.3) {\(a_1,a_4\)};
    \fill[orange!75!black] (240:1) circle (0.045);
    \node[font=\scriptsize] at (240:1.3) {\(\textcolor{orange!75!black}{a_2},a_5\)};
    \node[font=\small] at (0,-1.65) {\(h=2:\; e^{i2\theta_k}\)};
  \end{scope}
  
  \begin{scope}[shift={(7.2,0)}]
    \draw[black!40] (0,0) circle (1);
    \draw[->,black!35] (-1.25,0)--(1.25,0);
    \draw[->,black!35] (0,-1.25)--(0,1.25);
    \fill[orange!75!black] (0:1) circle (0.045);
    \node[font=\scriptsize] at (1.2,0.3) {\(a_0,\textcolor{orange!75!black}{a_2},a_4\)};
    \fill[blue!60!black] (180:1) circle (0.035);
    \node[font=\scriptsize] at (-1.2,0.3) {\(a_1,a_3,a_5\)};
    \node[font=\small] at (0,-1.65) {\(h=3:\; e^{i3\theta_k}\)};
  \end{scope}
\end{tikzpicture}
\caption{Cyclic encoding for period \(p=l=6\), at three harmonics.
Coefficient \(a_k\) is encoded at angle \(\theta_k = 2\pi k/p\). Harmonic \(h\) stores the
real and imaginary parts of \(e^{2\pi i\cdot k\cdot h/p}\).
Adjacent coefficients (modulo \(l,m\)) are mapped to adjacent angles. At \(h=1\) the map is one-to-one.
At \(h=2\) the map identifies slots \(k\) and \(k+3\). The harmonics \(h \le \lfloor p/2\rfloor = 3\) form a complete Fourier basis
for functions of position on the period-\(6\) cycle, which is why we chose \(H\geq 3\) for training on \(l=6=m\) codes.}\label{fig:unitcircle}
\end{figure*}

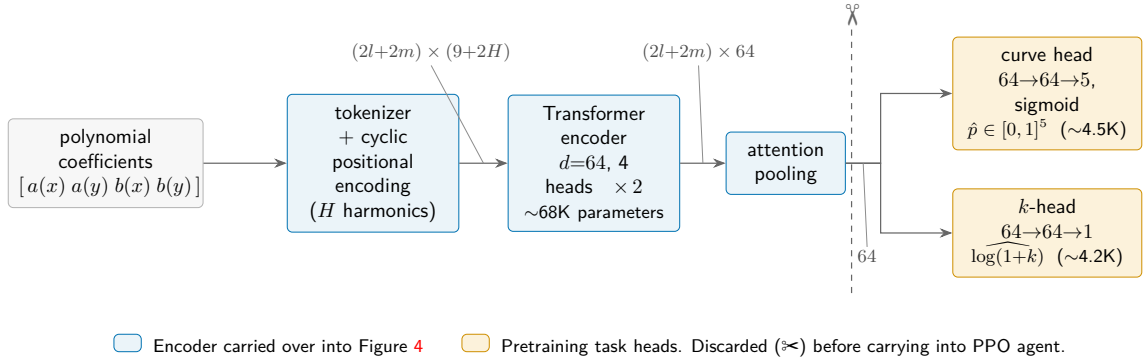
\begin{figure*}[htbp]
\centering
\resizebox{\linewidth}{!}{%
\begin{tikzpicture}[archfont, >={Stealth[length=2.2mm]}]

\node[input] (poly) at (0, 0) {polynomial coefficients\\ $[\,a(x)\; a(y)\; b(x)\; b(y)\,]$};
\node[pre] (tok)  at (4.6, 0) {tokenizer + cyclic\\ positional encoding\\ ($H$ harmonics)};
\node[pre] (trf)  at (8.4, 0) {Transformer encoder\\ $d{=}64$, 4 heads\quad $\times\,2$\\ {\footnotesize $\sim$68K parameters}};
\node[pre, text width=17mm] (pool) at (11.7, 0) {attention\\ pooling};

\node[head] (curve) at (16.2,  1.2) {curve head\\ $64{\to}64{\to}5$, sigmoid\\ {\footnotesize $\hat p\in[0,1]^5$ \;($\sim$4.5K)}};
\node[head] (khead) at (16.2, -1.2) {$k$-head\\ $64{\to}64{\to}1$\\ {\footnotesize $\widehat{\log(1{+}k)}$ \;($\sim$4.2K)}};

\draw[arr] (poly) -- (tok);
\draw[arr] (tok) -- coordinate[midway] (m1) (trf);
\draw[arr] (trf) -- coordinate[midway] (m2) (pool);
\draw[arr] (pool.east) -- ++(6mm,0) coordinate[midway] (m3) |- (curve.west);
\draw[arr] (pool.east) -- ++(6mm,0) |- (khead.west);

% Text callouts with diagonal lines, since when I rendered this the text was mushed up together
\node[callout] (c1) at (5.6, 1.9)  {$(2l{+}2m)\times(9{+}2H)$};
\draw[lead] (c1.south) -- (m1);
\node[callout] (c2) at (10.2, 1.9) {$(2l{+}2m)\times 64$};
\draw[lead] (c2.south) -- (m2);
\node[callout] (c3) at (13.1, -1.6) {$64$};
\draw[lead] (c3.north) -- (m3);

% cut line: the task heads are discarded after pretraining
\draw[dashed, black!60, semithick]
  ($(pool.east)+(1mm,2.3)$) -- ($(pool.east)+(1mm,-2.3)$);
\node[font=\large, rotate=-90, text=black!60] at ($(pool.east)+(1mm,2.55)$) {\ding{34}};

\node[anchor=north west, align=left, font=\sffamily\footnotesize, text=black]
  at (0, -2.8)
  {\tikz{\node[pre, text width=4mm, minimum height=3mm, inner sep=1pt]{};}\;
     Encoder carried over into Figure~\ref{fig:hybrid}\qquad
   \tikz{\node[head, text width=4mm, minimum height=3mm, inner sep=1pt]{};}\;
     Pretraining task heads. Discarded (\ding{34}) before carrying into PPO agent.};

\end{tikzpicture}%
}
\caption{The encoder trained in advance by the
multi-task loss of Section~\ref{sec:critic}. A shared trunk feeding a curve
head and a \(k\)-head. The parameters depend only on
\((H, d_{\mathrm{model}}, \text{heads}, \text{layers})\), but not on \((l,m)\), i.e., 
the token count \(2l+2m\) varies with code size, but the token width \(9+2H\) does
not. The two task heads are discarded once training is done. The rest of the network
(blue, \(\sim\)68K parameters at \(H{=}3\)) is carried over to the agent architecture in Figure~\ref{fig:hybrid}.}
\label{fig:encoderArchitecture}
\end{figure*}

\noindent The encoder architecture is given in Figure~\ref{fig:encoderArchitecture}

\subsection{The logical-qubit features\label{sec:kFeatures}}
Recall from Definition~\ref{def:reward} that if the number of logical qubits \(k\), is below \(k_{\min}\) 
the reward is a deterministic function of \(k\). The environment computes \(k\) by Gaussian elimination over \(GF(2)\) at every step.
It is therefore sensible to help both the agent and the critic to consume \(k\) as an input, and address it immediately. 
Otherwise, we would have to either teach the agent to perform Gaussian elimination, which is somewhat wasteful, or, 
force (by some design insight) to steer towards codes with sufficiently high number of logical qubits. We therefore pass three scalars,
\[
\big[\; \min(k-k_{min},0)/k_{min}, \quad \mathbf{1}_{k \ge k_{min}}, \quad \log(1+k) \;\big],
\]
i.e.:
\begin{enumerate}
\item The (normalised) difference between the minimum number of logical qubits required
\item An indicator of whether the requirement is met, and
\item \(\log(1+k)\)
\end{enumerate}

\subsection{The fusion layer\label{sec:fusion}}
The three branches are combined by concatenation followed by a single \(\tanh\) layer,
\[
  323 = 64_{\mathrm{enc}} + 256_{\mathrm{mlp}} + 3_{k} \;\longrightarrow\; 256 ,
\]
the output of which is the representation read by the instance head (two instances, one for the critic, one for the actor).

\noindent Fusing the three branches of the architecture later rather than sooner allows us to inspect 
each branch, and potentially hot-plugging branches between trained models, interchanging branches without harming the rest of the network.

\subsection{The policy head and the value head\label{sec:heads}}
The action as we defined it in Section~\ref{sec:mdp} is a tuple \(A^{a(x)},A^{b(x)},A^{a(y)},A^{b(y)}\) of four
per-polynomial actions, each of which flips at most one coefficient. This dictates what the policy head 
should look like. It is made of four independent categorical distributions, one per polynomial, 
of size \(l+1\) for \(a(x),b(x)\) and \(m+1\) for \(a(y),b(y)\), i.e., 
one entry per coefficient, and an additional ``no-op''.
This makes (\(7\) categories each for \(l=m=6\)). 
The four are conditionally independent given the state, 
so the log probability of a joint action is the sum of the four log probabilities. 
%and the entropy bonus decomposes in the same way.
\noindent We point out, that single cells changes of \(H^X,H^Z\) directly, using an independent Bernoulli output per matrix cell, 
the head would have \(\mathcal{O}({(lm)}^{2})\) outputs, quadratic in the code size.
This is a place to state again, that the policy construction benefits from mirroring the code construction, and that otherwise the agent would have to learn
(or otherwise be constrained) to reproduce code qualities, a sampling inefficiency.

\noindent The value head is a single linear map \(256 \to 1\) returning \(\hat V(s)\), trained on the PPO value
target of Algorithm~\ref{alg:main}.
Since the actor and the critic are separate instances of Figure~\ref{fig:hybrid}, 
the value head reads its own trunk, and the two networks share no weights during
PPO even though they may start from the same pretrained encoder. 
The reader should be aware that there is a body of work where the actor and critic share weights, and that we did not explore such 
a path, and leave it as potential further work.

\subsection{Pretraining the encoder\label{sec:critic}}
We now return to the branch of Section~\ref{sec:encoderBranch} and ask:\\
\textit{Can we train the encoder to improve sampling efficiency, and speed up training by guiding the agent to better codes ?}\\

\noindent Unlike the MLP branch, where the input width is a function of the code size, the input of the encoder has code-size-agnostic width.
The encoder maps a code to a latent description of that code.
This could be a supervised problem for which we already have data, i.e., the dataset of
Section~\ref{sec:dataset}, and Appendix~\ref{appendix:dataset}, which contains evaluated codes, each logged with its polynomials, its number
of logical qubits, and its Monte-Carlo error counts.

\noindent We therefore train the encoder in advance as a surrogate model, i.e., from a code's polynomials
\polynomials{} we train it to predicts
\begin{itemize}
      \item The monte-carlo combined error and failure rate curve and
      \item The number of logical qubits \(k\).
\end{itemize}
Which are the two ingredients that make up the reward in Definition~\ref{def:reward}.
These two tasks can be expressed as two terms of a loss function for the encoder, a multi-task sum
\begin{equation}
\begin{split}
  \mathcal{L} \;&=\; \underbrace{\tfrac{1}{5}\sum_{i=1}^{5} n\cdot\mathrm{BCE}\!\big(\hat p_i,\,c_i/n\big)}_{\text{Binomial NLL of the curve}}\\
  \;&+\; \lambda_k\,\underbrace{{\big(\widehat{\log(1+k)}-\log(1+k)\big)}^2}_{\text{\(k\)-head MSE}}
  \end{split}
\end{equation}
\noindent The two heads used here are discarded once pretraining is done, and the rest of the weights are carried over into
Figure~\ref{fig:hybrid} and plugged as the weights of the encoder that produced the \(64\)-dimensional latent representation.

\noindent The motivation behind this was, that a representation which is sufficient to predict how a code performs is also a
good starting point for a critic that has to estimate what the code value is, and for an actor that has to
decide how to improve it.

\noindent This extrapolation was expected to be hard, since for a fixed physical error probability \(p\), 
the probability that a sampled error is trivial is \({(1-3p)}^{N}\), where \(N=2lm\) is the number of physical qubits of a BB code. 
As \(N\) grows, this quantity reduces, and for a fixed number of samples in a monte carlo evaluation (we use \(50\) samples for each \(p\)) 
does not take this into account. In addition to this, a \(6,6\) code can have a maximum number of logical qubits of \(72\), and this number grows with the size of the code.

\noindent Our approach is to train the encoder at small \(l,m\) and then use it unchanged
or cheaply adapt (``ground'') it at larger sizes. The token design explained in Table~\ref{tab:token-features} of
Section~\ref{sec:encoderBranch} makes this possible in principle, i.e., features are
extracted identically at every size. Whether the learned weights actually succeed in extrapolating is a
question which we answer with a sweep over training recipes.
We trained over several \textit{\textbf{training recipes}}, which vary in which values of \((l,m)\in\{(6,6),\,(9,6),\,(12,6),\,(15,3)\}\) are we training on, 
how many harmonics are used \(H\), and which labels, the encoder consumes, as explained below:

\noindent \textbf{(6,6) only} using both labels: the error-rate data and the number of logical qubits \(k\).\\
\noindent\textbf{(6,6) + \(k\)-grounding} Fine-tune the above baseline at the three larger sizes
using the \(k\)-prediction task alone on the larger sizes, not using the error-rate data. Fine tuning is done using 2 epochs and at a lower training rate. The rationale is that \(k\) is
computable exactly at any size by Gaussian elimination over \(GF(2)\), so
this adaptation requires no Monte-Carlo decoder labels at larger sizes. We ground in two
variants, both include \(k\) loss (consuming the \(k\) labels of the bigger codes): \textbf{\textit{anchored}} and \textbf{\textit{non-anchored}}.
In the \textbf{anchored} variant, we revisit the original \(6,6\) curve task while training on new labels, as a guard against forgetting. In the \textbf{non-anchored} we do not revisit the \(6,6\).\\
\noindent\textbf{Mixed sizes} Train on mixed sizes jointly on the full task (curve and \(k\)) using the dataset codes in sizes:
\begin{itemize}
\item \((6,6)\) + \((9,6)\)
\item \((6,6)\) + \((9,6)\) + \((12,6)\)
\item \((6,6)\) + \((9,6)\) + \((12,6)\) + \((15,3)\)
\end{itemize}
\noindent Each recipe, as documented in Table~\ref{tab:recipes}, is trained several times with independent seeds.

\begin{table*}[t]
  \centering
  \caption{Encoder training recipes. Every recipe was trained at each
  \(H \in \{3,4,6,7,10\}\) with ten independent seeds (50 models per recipe,
  300 in total). Grounded models are initialised from the \((6,6)\)-only
  checkpoint with matching \(H\) and seed, so ground-vs-baseline comparisons
  are paired by lineage. Each size's pool is partitioned \(80/10/10\) into
  training, validation and test sets by the model's own seed; ``Training
  codes'' is the training partition summed over the recipe's sizes, and the
  best epoch is selected on the validation partition. Every trained model is
  then evaluated on the full pools of all seven code sizes which include codes that the model has seen.}
  \label{tab:recipes}
  \small
  \begin{tabular}{lllrrcc}
    \toprule
    Recipe &  \makecell{\(k\)\\labels} & \makecell{Curve\\labels}
    & \makecell{Training \\codes} & \makecell{Evaluation\\ codes} & lr & \makecell{Best epoch\\ (val)} \\
    \midrule
    \mixedSixSix                      & \((6,6)\) & \((6,6)\)
      &    716\,040 & 38\,005\,438 & \(10^{-3}\) & 23 [20--23] \\
    \mixedSixSixSixNine                  & \(+(9,6)\) & same as \(k\)
      & 2\,912\,631 & 38\,005\,438 & \(10^{-3}\) & 11 [10--11] \\
    \mixedSixSixSixNineOneTwoSix              & \(+\,(12,6)\) & same as \(k\)
      & 3\,995\,213 & 38\,005\,438 & \(10^{-3}\) & 7 [5--7] \\
    \mixedSixSixSixNineOneTwoSixOneFiveThree          & \(+\,(15,3)\) & same as \(k\)
      & 7\,525\,051 & 38\,005\,438 & \(10^{-3}\) & 5 [4--5] \\
    \midrule
    \groundAnchored &  all four & \((6,6)\) only
      & 7\,525\,051 & 38\,005\,438 & \(10^{-4}\) & 1 [0--1] \\
    \groundPure     &  all four & none
      & 7\,525\,051 & 38\,005\,438 & \(10^{-4}\) & 1 [1--1] \\
    \bottomrule
  \end{tabular}
\end{table*}

\subsubsection{Evaluation (of the encoder)\label{sec:encoderEvaluation}}
\begin{figure*}[htbp]
    \centering
    \begin{subfigure}{0.49\linewidth}
        \centering
        \includegraphics[width=\linewidth]{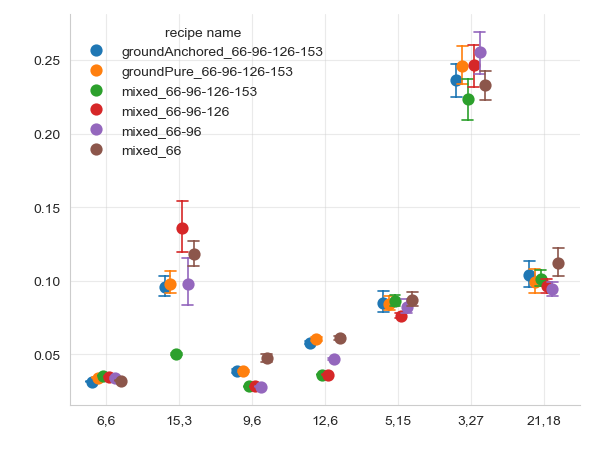}
    \end{subfigure}
    \hfill
    \begin{subfigure}{0.49\linewidth}
        \centering
        \includegraphics[width=\linewidth]{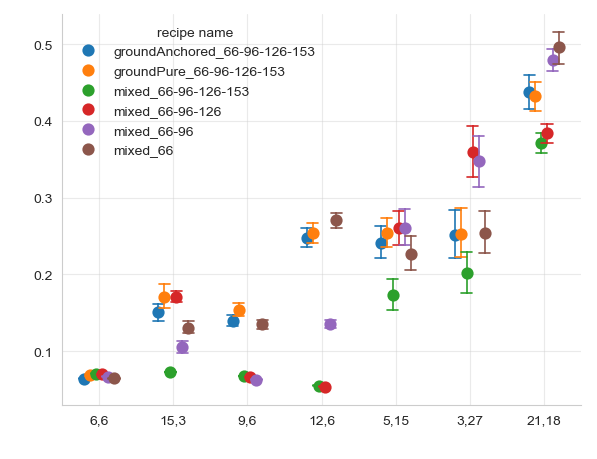}
    \end{subfigure}
    \caption{Mean absolute error of prediction vs. measured logical error rate across the \(300\) encoders trained, plotted for \(p\in\{0.01, 0.0316\}\).\label{fig:point_mae}}
\end{figure*}

\noindent We evaluate each model on the entire dataset, four sizes that were included in the training recipes, and three that were not, 
but using Mean Absolute Error (MAE):\\
\noindent \textbf{\(k\)-MAE} Error of the \(k\)-head in logical qubits. between actual number of logical qubits, shown in Figure~ ref{fig:eval-encoder-kMae}, and \\
\noindent \textbf{Curve-MAE} Per error point MAE of the predicted vs. measured shown in Figure~\ref{fig:point_mae}.\\
\noindent We note that this is not the metric on which training was done. While training was meant to enable transference, we evaluate the resulting models in MAE terms to see if transference was successful.

\noindent Table~\ref{tab:kCurveAlignment} shows a tension between successful prediction on \(k\) vs. the curve. 
By looking at rank correlation, we see that at unseen code sizes there is negative correlation, suggesting the encoder cannot succeed at extrapolating \textbf{both} metrics.
With respect to predicting the curve, Figure~\ref{fig:point_mae} shows the prediction for \(p\in\{0.01,0.0316\}\). As expected, predicting the error rate is harder a larger code sizes, but all models seem to be clustered (in agreement), including their in their collective failure at the \((3,27)\) size at \(p=0.01\) and \(21,18\) at \(p=0.0316\).

\begin{table*}[t]

  \centering
  \caption{Rank correlation between the encoder's \(k\)-MAE and its curve-MAE.
  Rows are training recipes, columns are the code parameters \((l,m)\) the encoder
  was evaluated on. A positive entry means the codes whose \(k\) is predicted well are
  also the codes whose error curve is predicted well. A negative value means the two
  objectives disagree. The three rightmost sizes appear in none of the recipes, so they
  are pure extrapolation. The negative correlations in the last row, where the model was
  trained on the \((6,6)\) codes dataset and grounded using \(k\) labels, as well as in
  the rightmost three columns, suggest there is a tension between predicting \(k\) and
  predicting the performance of a code.}
  \label{tab:kCurveAlignment}
  \footnotesize
  \setlength{\tabcolsep}{8pt}
  \renewcommand{\arraystretch}{1.15}
  \begin{tabular}{@{}l *{4}{S[table-format=-1.2]} @{\hspace{2em}} *{3}{S[table-format=-1.2]}@{}}
    \toprule
    %\multirow{2}{*}{\diagbox[width=7cm,height=2.7\line]{\textbf{Recipe}}{\textbf{Code parameters} \((l,m)\)}}
    & \multicolumn{4}{c}{Sizes named in the recipes}
    & \multicolumn{3}{c}{Unseen sizes} \\
    \cmidrule(lr){2-5} \cmidrule(l){6-8}
    & {\((6,6)\)} & {\((9,6)\)} & {\((12,6)\)} & {\((15,3)\)}
    & {\((5,15)\)} & {\((3,27)\)} & {\((21,18)\)} \\
    \midrule
    \texttt{\mixedSixSix}                     &  0.36 &  0.21 & -0.05 &  0.22 & -0.46 & -0.50 & -0.10 \\
    \texttt{\mixedSixSixSixNine}                  &  0.29 &  0.47 & -0.05 &  0.39 &  0.17 & -0.62 &  0.56 \\
    \texttt{\mixedSixSixSixNineOneTwoSix}              & -0.03 &  0.61 &  0.76 &  0.70 & -0.19 & -0.39 &  0.40 \\
    \texttt{\mixedSixSixSixNineOneTwoSixOneFiveThree}          &  0.03 &  0.12 &  0.34 &  0.39 & -0.53 & -0.40 &  0.06 \\
    \addlinespace
    \texttt{\groundAnchored} &  0.42 &  0.08 &  0.07 & -0.15 & -0.21 & -0.35 & -0.26 \\
    \texttt{\groundPure}     &  0.34 & -0.03 &  0.00 & -0.08 & -0.12 & -0.43 & -0.23 \\
    \bottomrule
  \end{tabular}
\end{table*}

% \begin{table}[t]
%   \centering
%   \caption{Each of the 300 encoders is evaluated on all
%   seven sizes. ``In recipes'' marks the four sizes that appear in at least one
%   training recipe. The table counts records that are distinct polynomials (recall that this does not imply different codes) on the \texttt{geometric5} error range.}
%   \label{tab:evaluationPools}
%   \begin{tabular}{lrrc}
%     \toprule
%     \((l,m)\) & \(n=2lm\) & Codes & In recipes \\
%     \midrule
%     \((6,6)\)   &  72 &    895\,050 & \checkmark \\
%     \((15,3)\)  &  90 &  4\,412\,297 & \checkmark \\
%     \((9,6)\)   & 108 &  2\,745\,740 & \checkmark \\
%     \((12,6)\)  & 144 &  1\,353\,227 & \checkmark \\
%     \((5,15)\)  & 150 & 18\,402\,947 & \\
%     \((3,27)\)  & 162 & 10\,148\,123 & \\
%     \((21,18)\) & 756 &     48\,054 & \\
%     \midrule
%     Total       &     & 38\,005\,438 & \\
%     \bottomrule
%   \end{tabular}
% \end{table}

\section{Related Work}\label{sec:related-work}
At the time of writing, the field of quantum error correction 
codes is experiencing a montage of \(70\) years worth of classical information theory, 
error correction codes and machine learning, making it hard to stay up to date.
In writing this section we aimed to give the reader a grounding in related QECC,
and specifically machine learning for ECC and QECC. 
Two resources were instrumental in laying the ground work in QECC for this work. 
The first is by Pantellev and Kalachev~\cite{panteleev2021degenerate}, 
which present BP+OSD decoding used here, as well as multiple code constructions, and methodology notes.
The second is Gottesman's book-in-writing~\cite{gottesman2026surviving}, 
which serves as a one-stop-shop for everything QECC related, 
and also makes a brilliant read\footnote{With pearls of wisdom such as ``...lattice surgery is an elective procedure...''}. 
The decoder implemented in this work is the one reported in~\cite{roffe_decoding_2020}, 
which is a CPU bound implementation. 
For the Belief Propagation step, specifically Min-Sum decoding, 
we considered using~\cite{sella2025mapping}, but have not yet paired it with the 
OSD post processing stage, which is needed to avoid CPU-GPU contention. 
For the BB family specifically, we note that search algorithms were reported by~\cite{wang2026coprime, wang2024coprime}.
\noindent Parametrised constructions of quantum Low Density Parity Check (qLDPC) Error Correcting Codes (ECC) 
draw from several mathematical constructs like polynomials~\cite{panteleev2021degenerate}
and graphs~\cite{evra2022decodable, manjunath2026universal}, 
but ultimately require an assignment of these parameters for practical use, 
and~\cite{panteleev2021degenerate} raises the question of finding good parameters. 

\noindent Reinforcement learning has been applied to the actual decoding as shown in~\cite{sweke2020reinforcement, blue2026machine}, as well as
to modify surface-code construction in~\cite{nautrup2019optimizing}, 
optimizing tensor-network codes as reported in~\cite{mauron2024optimization}, 
and reducing stabiliser weights of existing qLDPC codes~\cite{he2025discovering}. 
These works reward primarily structural properties of code parameters. 
In contrast, our main focus is the measured performance of the code under a fixed decoder.
Machine learning alternatives to RL have appeared recently as well, one workflow that uses LLM was introduced in~\cite{cruzbenito2026evolutionary}, and, closer to our \emph{encoder} approach, 
a Bayesian-optimisation framework that trains a model that predicts logical error rates without decoder simulation, 
and uses it to discover competitive \([[144,36]]\) and \([[144,16]]\) codes~\cite{cheng2026bayesian}. 
In our work this type of model is deployed to warm-start an RL agent, which trains a generative policy. 
Conceptually, the closest work to ours is~\cite{su2305discovery}, which sets to discover QECCs with \(k=1\) logical qubits using an analytic proxy physical-to-logical error rate. 
The work in~\cite{su2305discovery} applies RL methods to Quantum Lego~\cite{cao2022quantum}, a way to build QECCs from tensor networks.
Finally, this work is continuation of our previous work on co-designing (classical) LDPC codes and decoders~\cite{sella2024coding}. 

\section{Evaluation and results}\label{sec:evaluation}
We evaluate the policy every \(10\) collector batches in a deterministic modes, by taking the mode of the the policy distribution. 
%Evaluation may have a different rollout length, but in any case (\(\geq \max(l,m)\)). 
The evaluation starts at random polynomials with up to \(3\) non zero elements (similarly to training). 
Every step of every evaluation is logged, giving, per evaluation, the full trajectory of rewards and of the constructed codes. 
The following \emph{metrics} are observed:
\paragraph{Best-performing code} We are interested, of course, in maximising the reward under a given decoder. For known, published, code parameters, we can calculate this number and measure against it.
\paragraph{80\% of the evaluation} This is of interest because it helps understand whether the agent is actually improving, in which case the 80\% band will tighten in width, and tighten the gap with the maximum. 
A persistent gap between the band and the maximum means the best
codes visited are fewer than $10\%$ of the time, implying that discovered good codes are not integrated into the policy. 
A persistently wide 80\% band is a proxy metric to 
\textit{how many steps did it take the agent to reach the best code ?} 
as an all knowing agent would plot a reward-greedy course towards the best code. The best code (the reference published code) is at most \(max{l,m}\) actions far, 
but potentially goes through some bad-paying codes. This is why we look at the fraction of steps the agent spent in \(5\) regions:
\begin{enumerate}[label = \roman*]
        \label{list:regions}
\item \textbf{Invalid codes}
\item Codes that yielded \textbf{0-50\%} of the reference reward
\item Codes that yielded \textbf{50-95\%} of the reference reward
\item Codes that yielded \textbf{95-100\%} of the reference reward
\item Codes that yielded a reward equal to or greater than the reference.
\end{enumerate}

\paragraph{Average reward} The average reward can indicate whether or not the agent has learned that codes with insufficient number of logical qubits obtain a negative reward.
\paragraph{Within-episode gain} This is the difference between the the reward for the code the envornoment \bfit{reset} into, and the \bfit{last} code the agent arrived in. Where learning was succesful, we expect this value to increase over time. 
This metric helps to filter out any episodes where the agent received a ``lucky'' reset value, and 

\begin{figure*}[htbp]
    \centering
    \includegraphics[width=\linewidth]{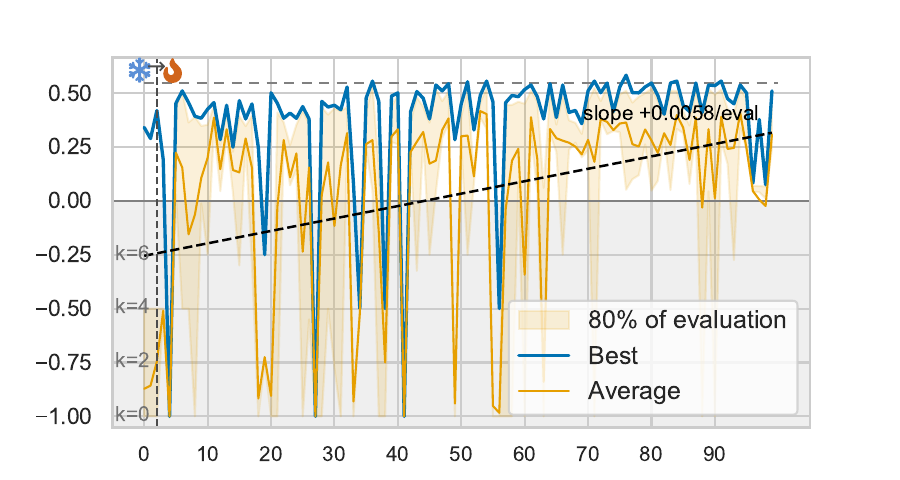}
    \caption{Evaluation of a single PPO training session. Horizontal dashed line marks the reward for a published reference code (which is expected to be the maximum possible reward). Vertical dashed line marks when the encoder weights were unfrozen \faSnowflake{} $\rightarrow$ \faFire{}. As learning progresses, the 80\% band (transparent orange) is contracts towards the best codes in the evaluation and lifts off from the penalty region (shaded lower part).}
    \label{fig:reward-panel}
\end{figure*}

\noindent Figure~\ref{fig:reward-panel} summarises several metrics, carried over \bfit{one} training run. We use the codes reported in~\cite{bravyi2024high} as 
baselines. Where a baseline code exists, we plot the reward this code would achieve as a reference (horizontal, gray, dashed). The shaded region below zero reward marks where the reward was negative due to a code with less than the minimum number of logical qubits required, 
and the exact number of logical qubits can be recovered from the negative reward (annotated next to the reward ticks). A dashed vertical line indicates when the encoder weights were unfrozen. Within each evaluation we plot the \bfit{best} (maximum) reward obtained (blue), the \bfit{average} (orange), 
and the 10th to 90th percentile of the rewards obtained in that evaluation in shaded orange\footnote{Which will always be around the average and below the maximum.}, which indicates where 80\% of that evaluation was spent.

\paragraph{Anatomy of a single run}
Early in training the 80\% band is high-spread and sits largely inside the penalty region (shaded gray) and the agent spends most of its steps climbing from the random reset 
code towards codes with more logical qubits. 
As training progresses, a successful run shows the band shifting away from the penalty region. 
As training matures, we see the best-in-evaluation line (blue) crossing the reference more consistently. 
We also see that the 80\% band contracts later in training. A deep dip accompanied by a high-spread band is a
slow recovery, suggesting the evaluation started from an unfavourable and the agent was slow to recover. 
A dip whose band is tightly centered around $-1$ suggests a trapped evaluation, i.e., the deterministic policy cannot climb out of the penalty region.
Recall that evaluations are deterministic. A trapped evaluation reflects a property
of the greedy policy from that particular initial code. The training time
policy, which samples with an entropy bonus, wouldn't necessarily fail from the same state, but such an event shows a weakness of the deterministic policy. 
These observations can be seen in Figure~\ref{fig:reward-panel}.

\begin{figure*}
        \centering
        \begin{subfigure}{0.48\linewidth}
        \includegraphics[width=\linewidth]{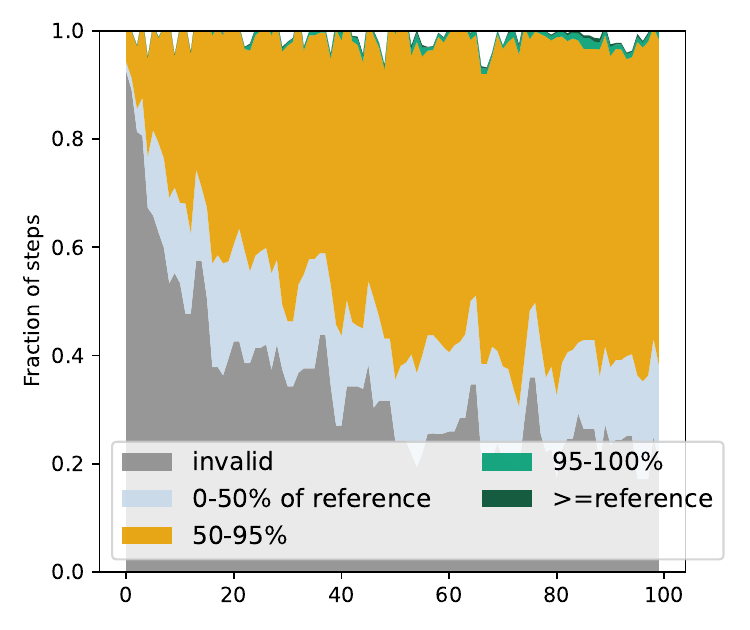}
        \end{subfigure}
        \hfill
        \begin{subfigure}{0.48\textwidth}
        \includegraphics[width=\linewidth]{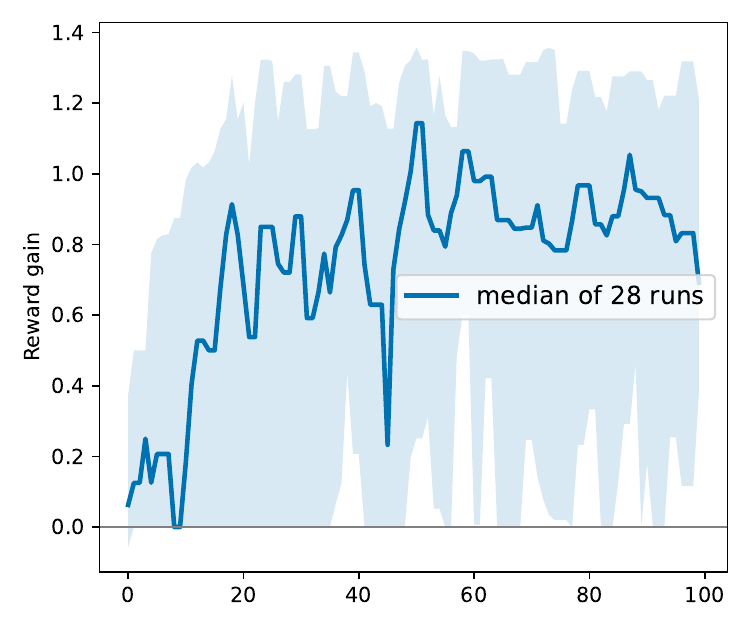}

        \end{subfigure}
                \caption{Left: stacked plot corresponding to the regions defined in \ref{list:regions}. The reference code is considered the best possible. As training progresses, the agent spends more of the evaluation time in better performing regions. 
                Right: rolling median (pooled over one adjacent evaluation on either side) within-episode gain, calculated as the difference between the reward for the first and last code of the episode.}
        \label{fig:multiple96}
\end{figure*}

\paragraph{Analysis of multiple runs}
Figure~\ref{fig:multiple96} (left) show a stacked plot of where the agent spends evaluation time divided by the 
regions~\ref{list:regions}, for \(28\) runs for the \(9,6\) parameters (multiple seeds, multiple encoders). 
As training progresses, the agent spends less time on codes that have less the minimum required number of logical qubits.
\noindent Figure~\ref{fig:multiple96} (right) shows the median episode gain for the same \(28\) runs of the \(9,6\) 
code, with a 50\% band. As learning progresses, the median gain within episode increases up to evaluation \(50\), where it begins to stabilise. The reward for the reference code can be found in Table~\ref{tab:configuration}, \(0.546\) for the \(9,6\) code.

\section{Discussion and further work\label{sec:conclusion}}
We have presented a reinforcement learning framework for designing quantum error correcting codes as a function of decoder architecture, by formulating the problem as a sequential decision making problem.
The use of Reinforcement Learning tools such as PPO is made applicable by defining a reward function based on the area under the physical-to-logical-error-rate curve.
The learned policy can be used to edit, and potentially optimise existing codes. 

\subsection{Immediate further work}
By immediate we mean that an improvement of the existing codes could provide an answer.
\paragraph{Decoder budget} Follow up work could limit the episode on decoding budget similar to~\cite{pardo2018time}.
In our setting the natural limit is a budget of decoder invocations, since a Monte-Carlo simulation using a BP+OSD decoder is the main cost of a rollout.
\paragraph{Reset archive} In our work, each episode \bfit{reset} uses random polynomials (with up to 3 non zero coefficients), since 
a \bfit{reset distribution} that better covers the state space is known to positively impact policy quality~\cite{kakade2002approximately}. 
An immediate follow up is to use an archive of specific codes, 
which are to be remembered and explored repeatedly~\cite{ecoffet2021first}. This would entail taking a subset of the existing dataset (say of the worst codes) and retraining with reset codes sampled from this subset.
\paragraph{Decoder architecture, Error model}We did not consider OSD-w type postprocessing, i.e., allowing the decoder to scan for error patterns up to w coordinates away, nor did we investigate training with one decoder and evaluating with another. Another dimension which we consider immediate follow up work, is steering away from the symmetric probability in the depolarizing channel (the reader may recall that in section~\ref{sec:background} we noted that BB codes provide symmetric X and Z generators).

\subsection{Wider application}
The most natural extension of this work would be to frame other code constructions as a sequential decision making problem, i.e., an environment for a PPO agent to interact with.
The evaluation in this work uses a simplistic code-capacity noise model (independent depolarising noise on each physical qubit). 
A natural extension is circuit-level noise, where ancilla preparation, syndrome measurement, and gate errors are modelled explicitly~\cite{fowler2012surface}, and an RL agent could be rewarded for performance under a circuit-level simulator. 
We are also interested in expanding this work to lattice surgery~\cite{litinski2019game}, and optimise codes as a function of the complexity of lattice surgery on them.
At the time of submission, an interesting potential application came up following~\cite{manjunath2026universal}, in which context, the agent might have to learn automorphism groups to provide logical gates.\\
\noindent There is a very natural extension of this work to truly co-design codes and decoders: 
In~\cite{muller2025improved} the authors propose a parametrized decoder, made of several ``legs'' of Belief Propagation. By treating these parameters as actionable, just as we address the code parameters, one could optimize both decoder and code at the same time.

\noindent Finally, we wish to offer some self-criticism about the (non) necessity of this work. 
At the time of writing there are \(\mathcal{O}(1,000)\) distinct entities engaged in building quantum hardware for useful quantum computation. 
Assume that each entity will develop around \(\mathcal{O}(100)\) quantum devices, each with its own decoder architecture, 
and that neither will share these codes with the others. 
Optimistically, that's still only around \(\mathcal{O}(100,000)\) potential applications of this work as it stands.
Where we see this work going, and where we believe the use case could truly benefit from these methods, is when an arbitrary quantum algorithm is to be compiled to, 
(potentially) a Linear Combination of quantum Channels (LCC), that run on one of several hardware backends 
(which, at the time of writing, seem to only grow), 
each running their own set of potential QECC and instruction sets, 
and where each instruction (apply a gate, measure a qubit, extract syndromes, correct for errors) 
is attached to some latency cost and error rate which cascades. 
Once compiled, the user applies a \bfit{sequence of changes}. The question we ask in this case is:\\
\bfit{Can a generative policy be trained to optimise the compiled algorithm, in lockstep with changes to the algorithm ?}

\subsection{Reproducibility}
\noindent The dataset of BB codes used in this work is available to clone on \url{https://github.com/Omer-Sella/bbCodesDataset} or Zenodo~\cite{sella2026bbcodesdataset}. The repository includes a minimal script to reproduce the figures relating to statistical properties like the distribution of the number of logical qubits.\\
\noindent The model weights (checkpoints) for the encoder of Section~\ref{sec:encoderEvaluation} can be found at \url{https://github.com/Omer-Sella/bbCodeSurrogates} as well as 
Zenodo~\cite{sella2026bbcodesurrogates}, 
and include the script to produce Figures~\ref{fig:point_mae}. To evaluate the (\(300\)) models over the dataset, we advise cloning both repositories, adding an environment variable that includes the path to the dataset, and running the evaluation module in the encoder repository.
%The code used to train these models can be found here []. 
%A docker file that would bring up the environment used to traing the encoder can be found here [], however, be aware it clones the entire BB codes repository (~6GB). The environment (bb-gym), along with the code used for training in this paper can be cloned here []. The implementation of bb-gym roughly follows ???, while the code to train PPO is almost a drop-in-place reuse of torchrl tools ???.
\paragraph{Experimental configuration}
All reported experiments use the single-coefficient action space
(\texttt{--env-bit-flipping True}), episodes that reset to a random code with
at most three non-zero coefficients per polynomial
(\texttt{--env-reset-type random3}) and truncate after \(T=30\) steps, the
five-point \texttt{geometric5} grid, \(50\) Monte-Carlo samples and \(50\)
belief-propagation iterations per grid point, and the width-normalised reward
of Definition~\ref{def:reward} (\texttt{--env-reward-engineering True}).
Evaluations use deterministic action selection with a rollout length of \(30-60\)
steps, every tenth collector batch. 

\begin{table*}[t]
  \centering
  \caption{Per-size experimental configuration. \(A\) and \(B\) are \(lm \times lm\)
  binary circulant sums, so a code has \(n = 2lm\) physical qubits and the flattened
  code observation \(A \Vert B\) has \(2(lm)^2\) entries. The parameter \(k_{\min}\) is set
  to the logical dimension of the published reference code at that
  size~\cite{bravyi2024high}. Reference rewards are
  computed by the reward of Definition~\ref{def:reward} on the \texttt{geometric5}
  error range, with width normalisation.}
  \label{tab:configuration}
  \begin{tabular}{rrrrlrr}
    \toprule
    \(l\) & \(m\) & \(n = 2lm\) & \(A, B\) & Reference code & \(k_{\min}\) & Reference reward \\
    \midrule
     6 &  6 &  72 & \(36\times36\)   & \([[72,12,6]]\)       & 12 & 0.346 \\
    15 &  3 &  90 & \(45\times45\)   & \([[90,8,10]]\)       &  8 & 0.506 \\
     9 &  6 & 108 & \(54\times54\)   & \([[108,8,10]]\)      &  8 & 0.546 \\
    12 &  6 & 144 & \(72\times72\)   & \([[144,12,12]]\)     & 12 & 0.554 \\
     5 & 15 & 150 & \(75\times75\)   & ---                   & ---& --- \\
     3 & 27 & 162 & \(81\times81\)   & ---                   & ---& --- \\
    21 & 18 & 756 & \(378\times378\) & \([[756,16,\leq34]]\) & 16 & 0.600 \\
    \bottomrule
  \end{tabular}
\end{table*}

\begin{acknowledgements}
We thank Dr.\ Pavel Panteleev for helpful clarification and for guidance on methodology
in the early stages of this work.
\end{acknowledgements}

\appendix
\section{Dataset}
\label{appendix:dataset}
For each set of code parameters \(l,m\) we plot the distribution of the number of logical 
qubits which can be computed as explained in Equation~\ref{eq:codeDimension}, as well as a heatmap, that shows how many of the codes have that coefficient set to \(1\).
Finally, for each set of parameters we plot the logical error rate as a function of the physical error rate, averaged across all code with \(100\%\) variance. 
The latter should provide an idea of how a code chosen in random in that space would perform.

\begin{table*}[h]
  \centering
  \caption{Code-evaluation records collected per code size and physical-error grid
    (as of 27/07/2026). The original dataset was collected only on 
    a linear five-point grid: \(p \in \mathrm{linspace}(10^{-4}, 10^{-1}, 5)\).
    The union grid is the a 15-point superset that contains the linear grid together with 
    every geometric grid used in this work, so that one collection run serves all of them. 
    Geometric evaluations are obtained by filtering out the corresponding evaluation points 
    in the union grid records.}
  \label{tab:recordCensus}
  \begin{tabular}{lrrrr}
    \toprule
    \((l,m)\) & \(n = 2lm\) & Linear (5-pt) & Union (15-pt) & Total \\
    \midrule
    \href{https://github.com/Omer-Sella/bbCodesDataset/tree/master/l_6_m_6}{\((6,6)\)}   & 72  & 1\,948\,449 &    964\,343 &  2\,912\,792 \\
    \href{https://github.com/Omer-Sella/bbCodesDataset/tree/master/l_15_m_3}{\((15,3)\)}  & 90  &    458\,638 & 3\,850\,656 &  4\,309\,294 \\
    \href{https://github.com/Omer-Sella/bbCodesDataset/tree/master/l_9_m_6}{\((9,6)\)}   & 108 & 2\,681\,585 & 2\,257\,087 &  4\,938\,672 \\
    \href{https://github.com/Omer-Sella/bbCodesDataset/tree/master/l_12_m_6}{\((12,6)\)}  & 144 &    457\,979 & 1\,108\,666 &  1\,566\,645 \\
    \href{https://github.com/Omer-Sella/bbCodesDataset/tree/master/l_5_m_15}{\((5,15)\)}  & 150 &           0 & 16\,640\,753 & 16\,640\,753 \\
    \href{https://github.com/Omer-Sella/bbCodesDataset/tree/master/l_3_m_27}{\((3,27)\)}  & 162 &           0 & 9\,078\,743 &  9\,078\,743 \\
    \href{https://github.com/Omer-Sella/bbCodesDataset/tree/master/l_21_m_18}{\((21,18)\)} & 756 &        385 &      6\,726 &       7\,111 \\
    \midrule
    Total       &     & 5\,547\,036 & 33\,906\,974 & 39\,454\,010 \\
    \bottomrule
  \end{tabular}
\end{table*}

\begin{figure*}[htbp]
    \centering
    \begin{subfigure}{0.45\linewidth}
        \centering
        \includegraphics[width=\linewidth]{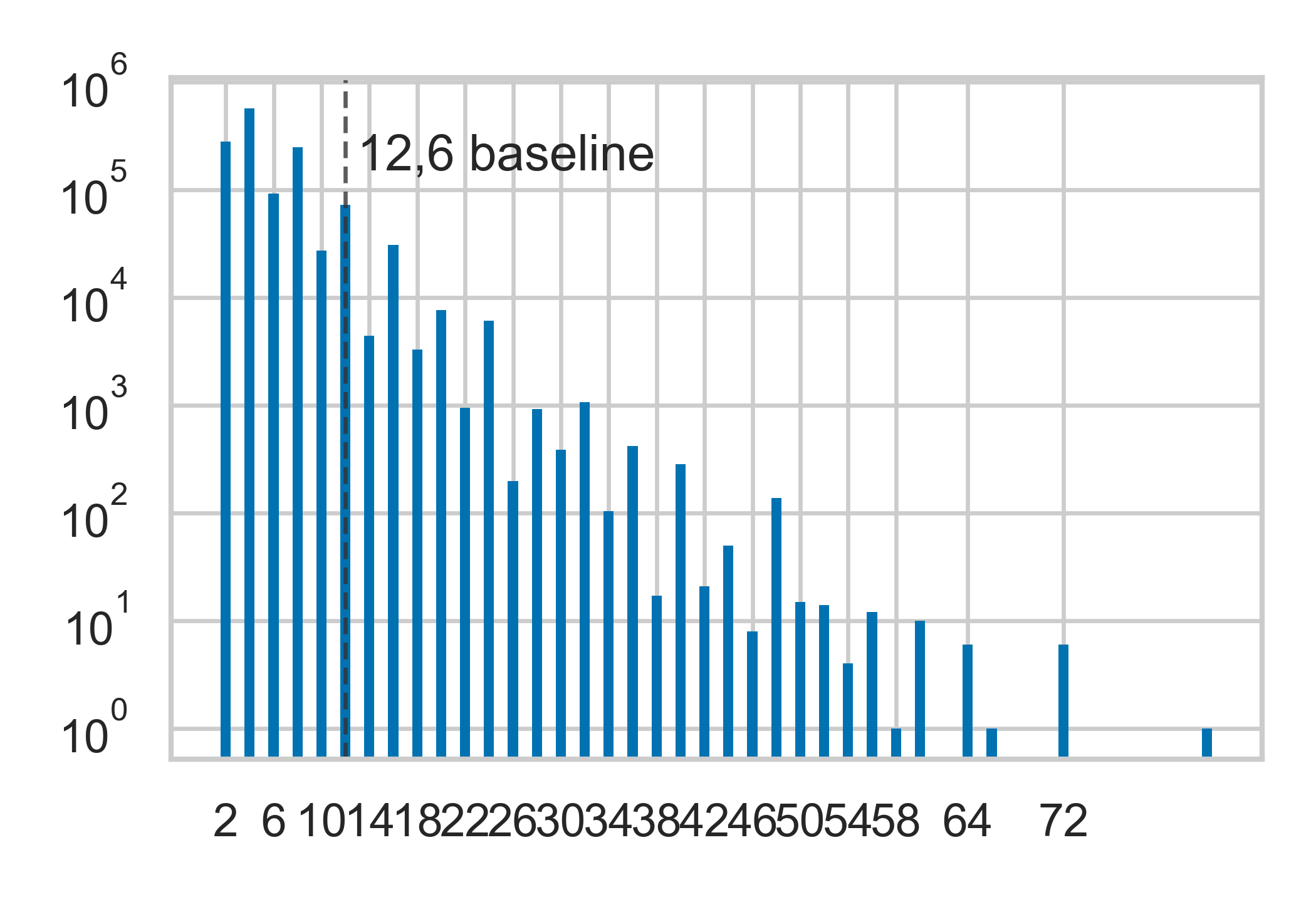}
        \caption{12,6}
    \end{subfigure}
    \hfill
    \begin{subfigure}{0.45\linewidth}
        \centering
        \includegraphics[width=\linewidth]{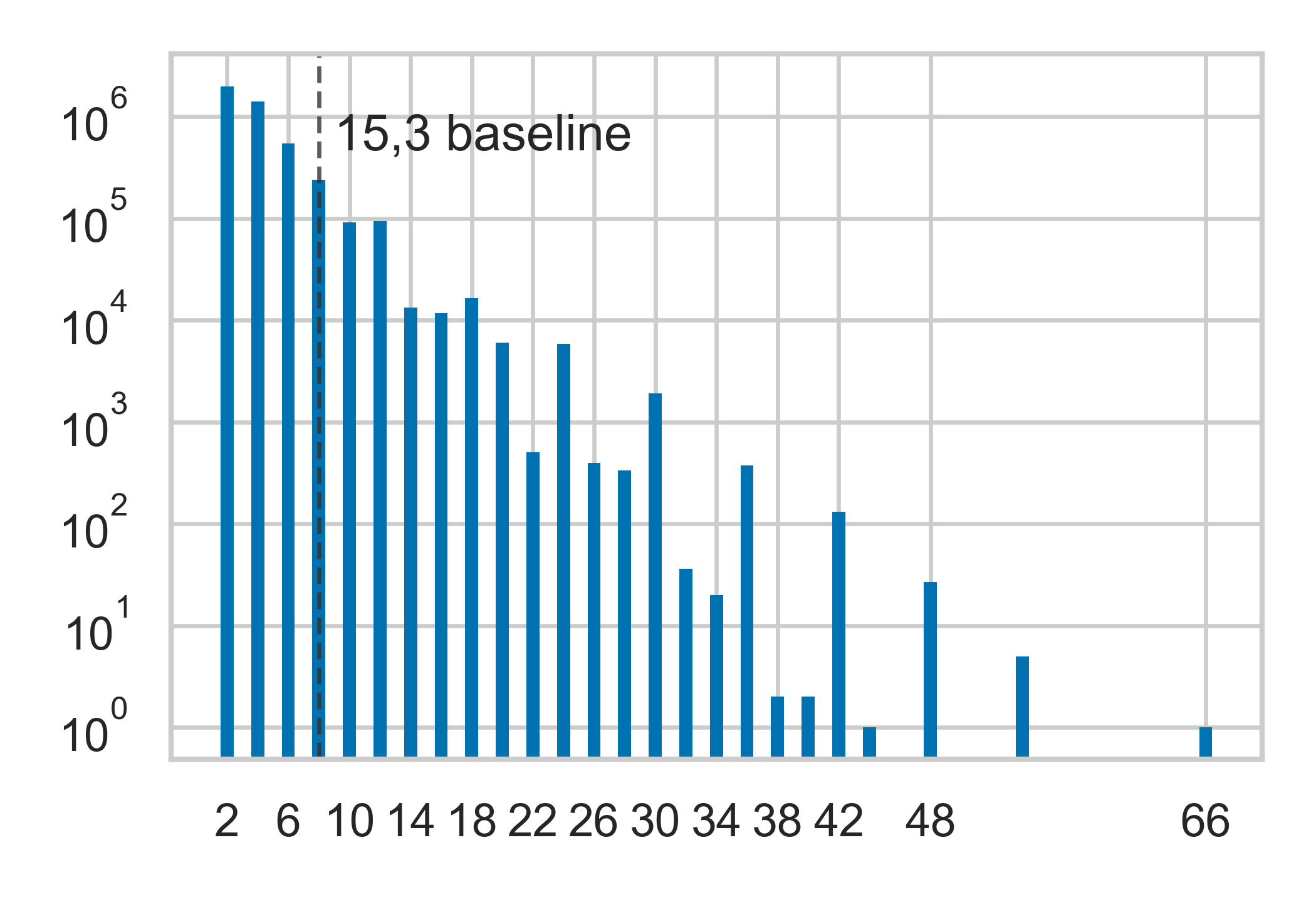}
        \caption{15,3}
    \end{subfigure}
    \vfill
    \centering
    \begin{subfigure}{0.45\linewidth}
        \centering
        \includegraphics[width=\linewidth]{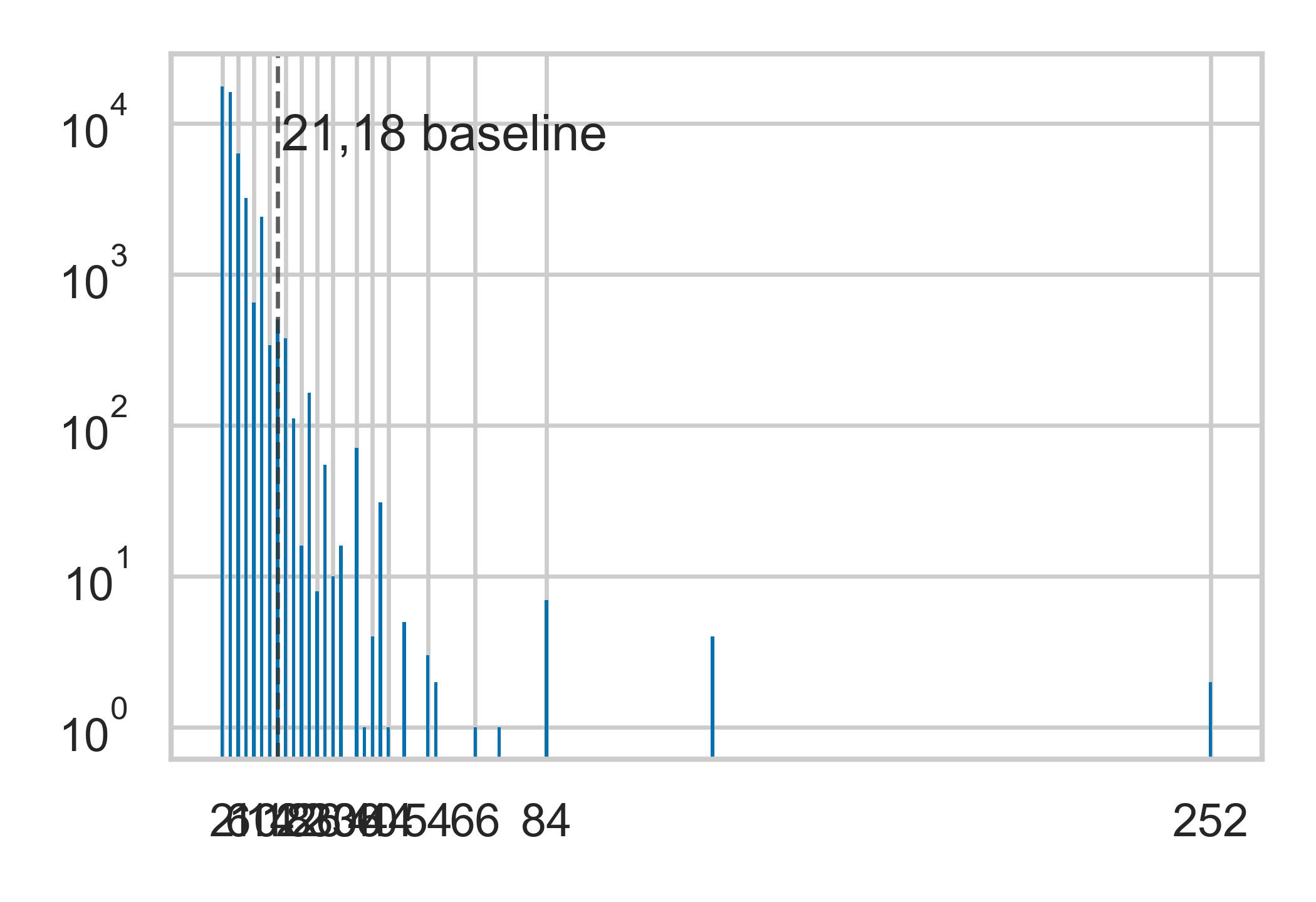}
        \caption{21,18}
    \end{subfigure}
    \hfill
    \begin{subfigure}{0.45\linewidth}
        \centering
        \includegraphics[width=\linewidth]{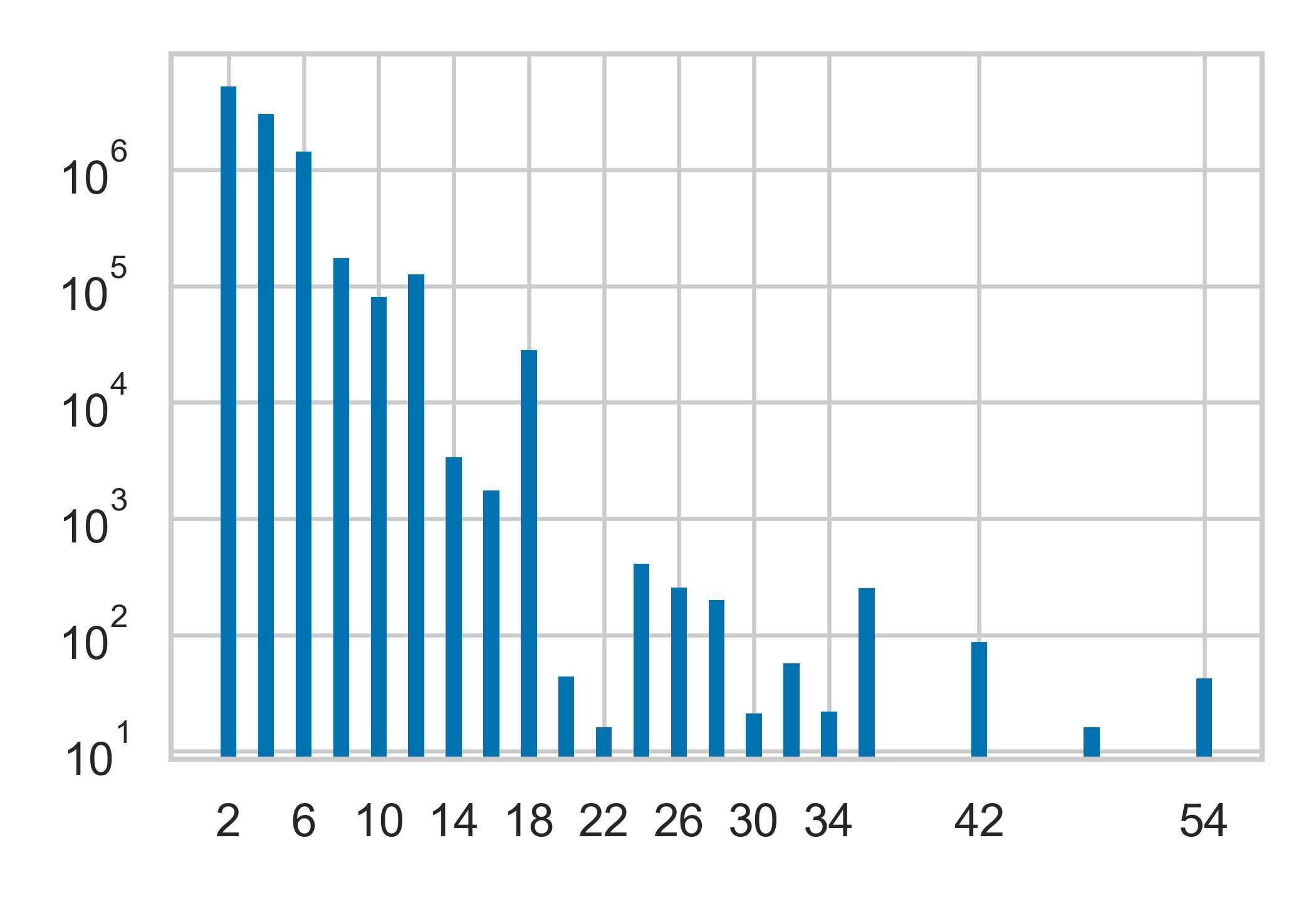}
        \caption{3,27}
    \end{subfigure}

    \vfill
    \centering
    \begin{subfigure}{0.45\linewidth}
        \centering
        \includegraphics[width=\linewidth]{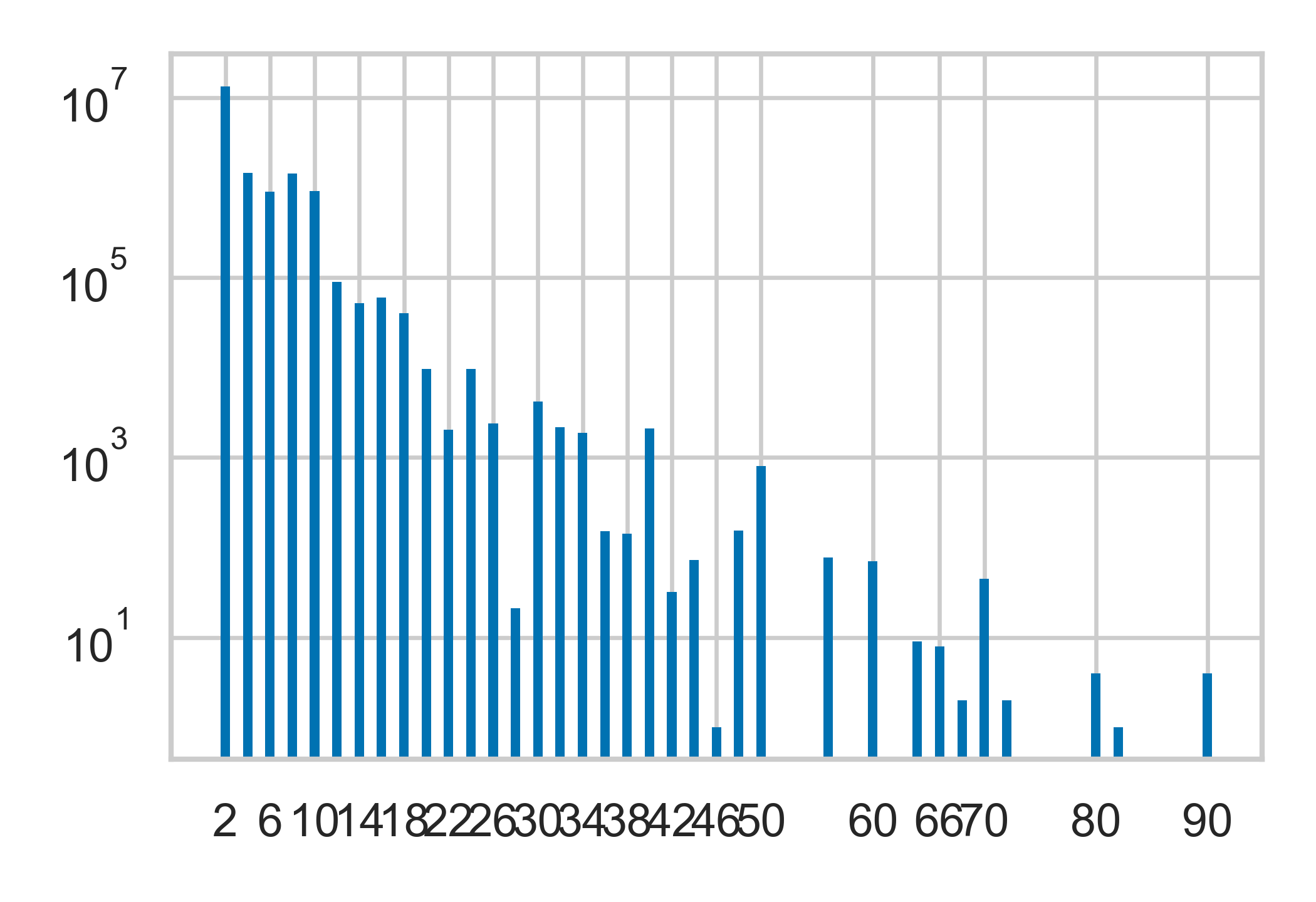}
        \caption{5,15}
    \end{subfigure}
    \hfill
    \begin{subfigure}{0.45\linewidth}
        \centering
        \includegraphics[width=\linewidth]{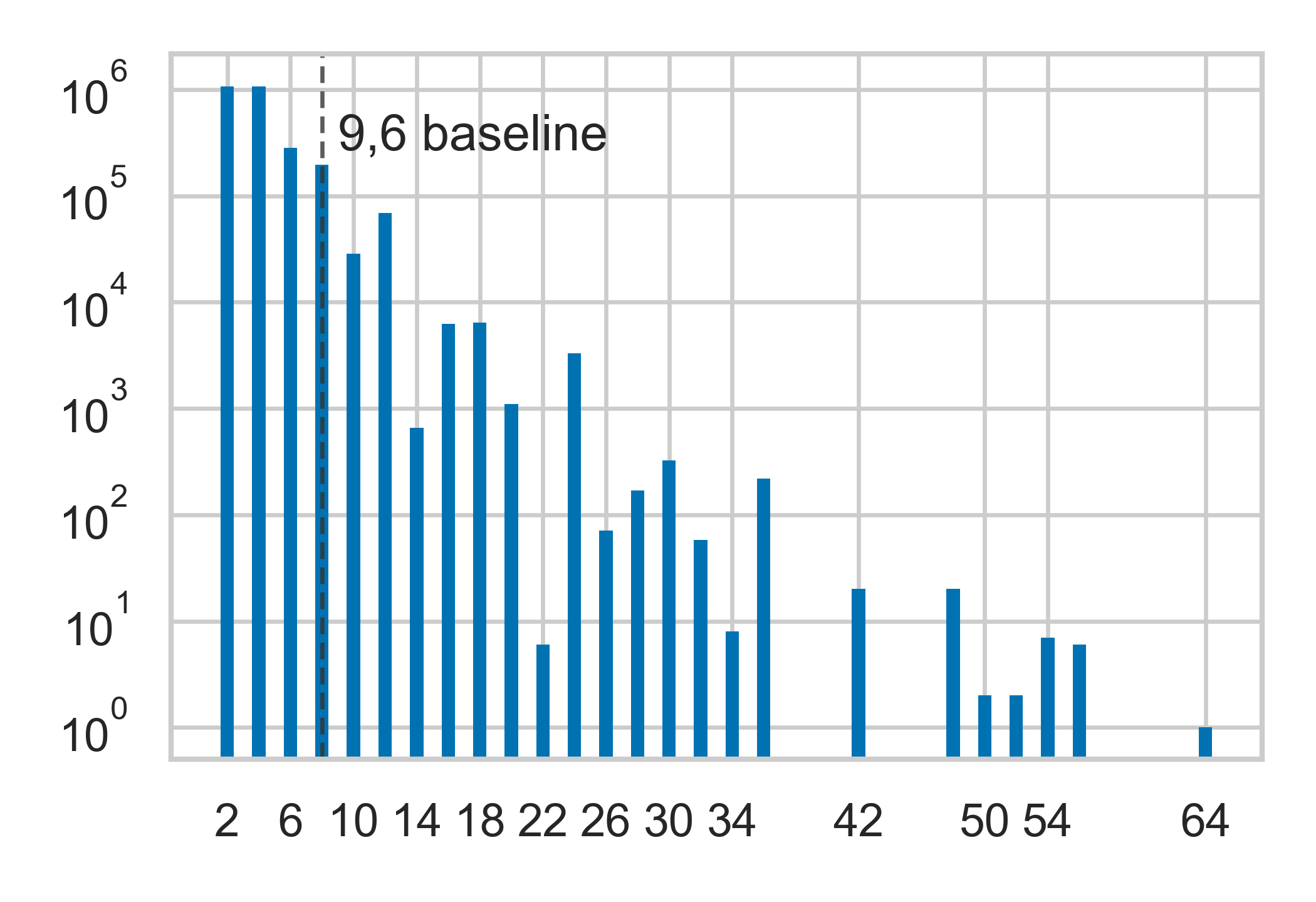}
        \caption{9,6}
    \end{subfigure}
    \caption{Distribution of the number of logical qubits over the dataset.}
    \label{fig:kDistribution-rest}
\end{figure*}

\begin{figure*}[htbp]
    \centering
    \begin{subfigure}{0.45\linewidth}
        \centering
        \includegraphics[width=\linewidth]{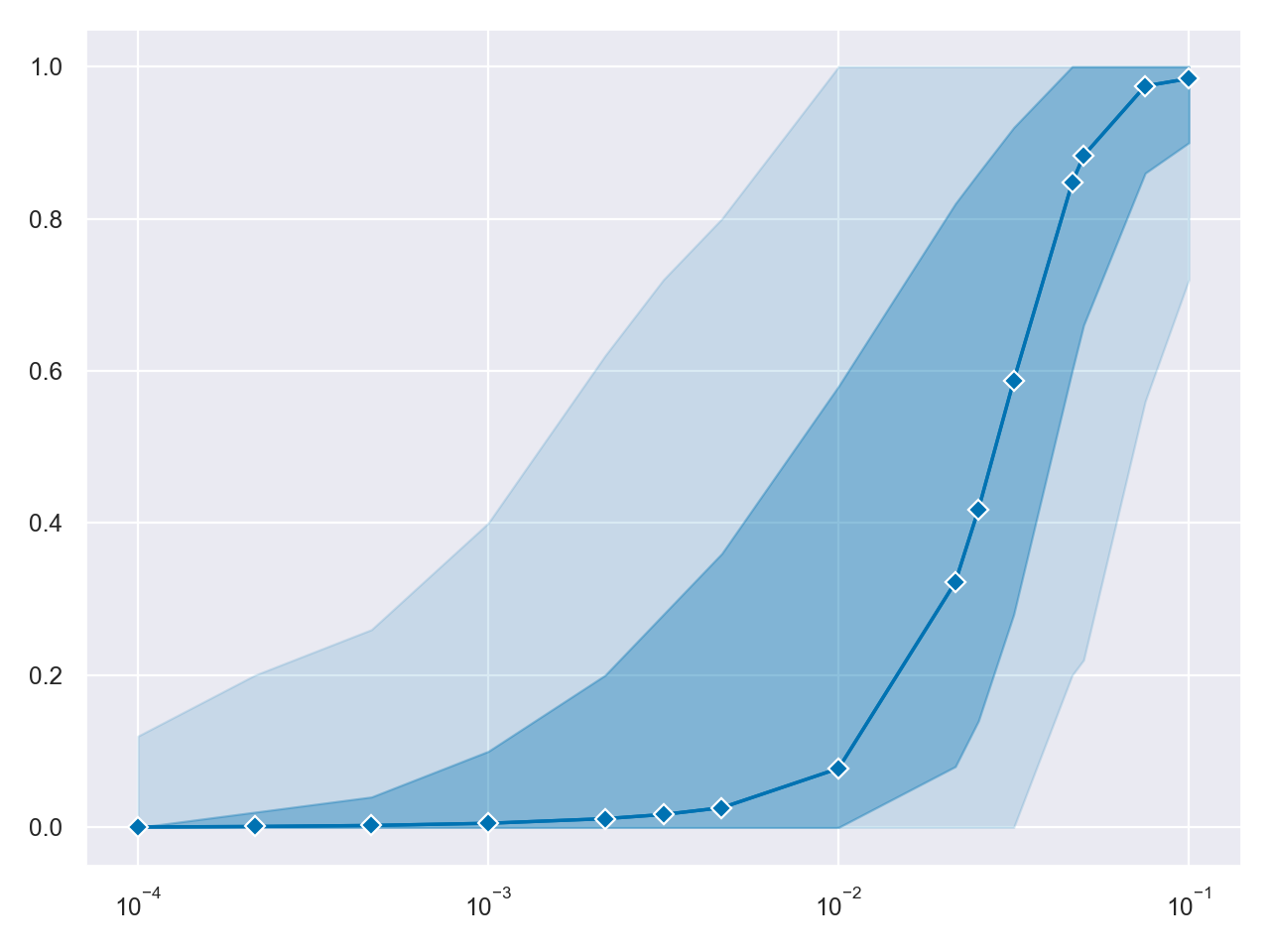}
        \caption{6,6}
    \end{subfigure}
    \hfill
    \begin{subfigure}{0.45\linewidth}
        \centering
        \includegraphics[width=\linewidth]{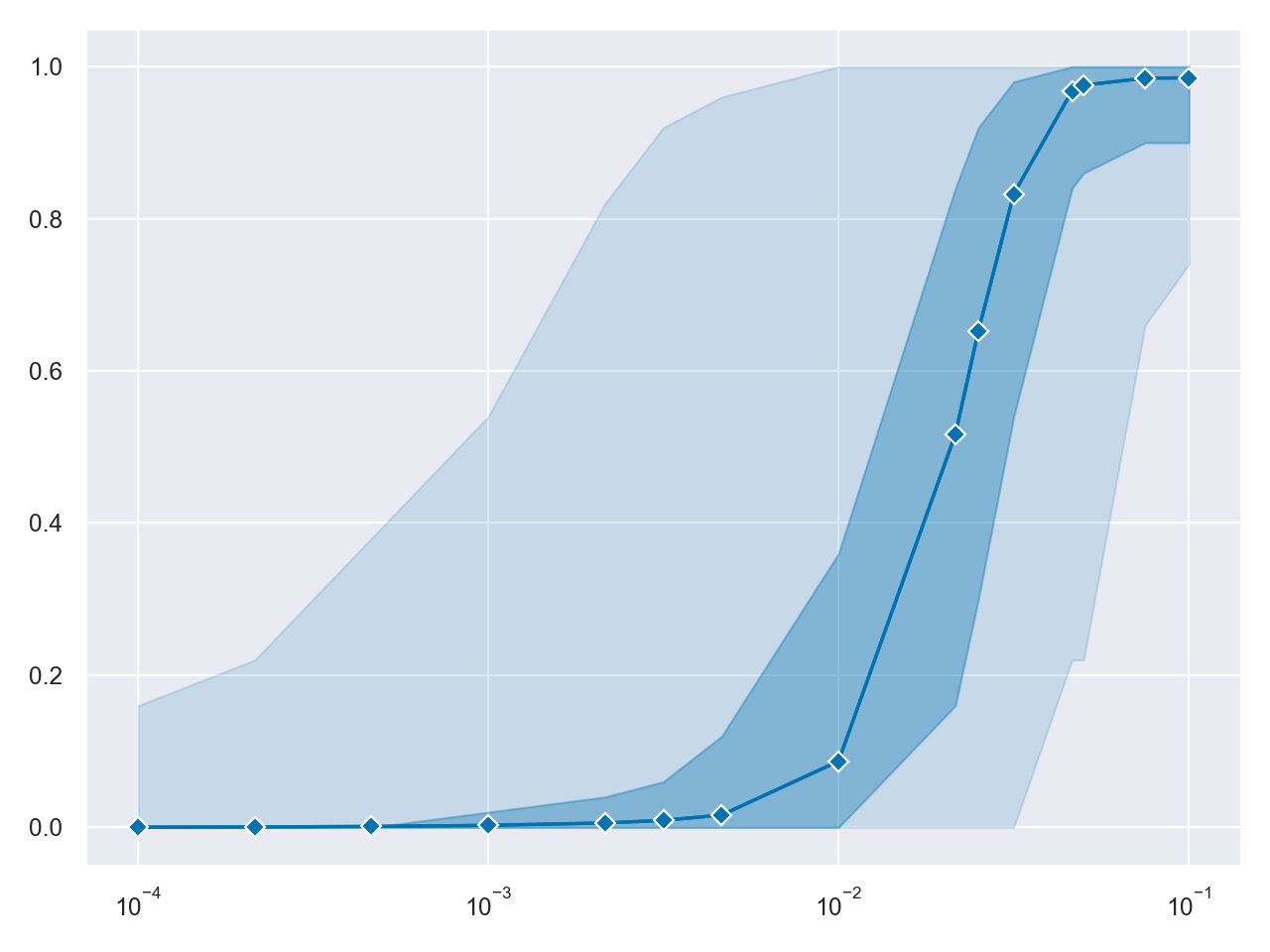}
        \caption{12,6}
    \end{subfigure}
    \vfill
    \begin{subfigure}{0.45\linewidth}
        \centering
        \includegraphics[width=\linewidth]{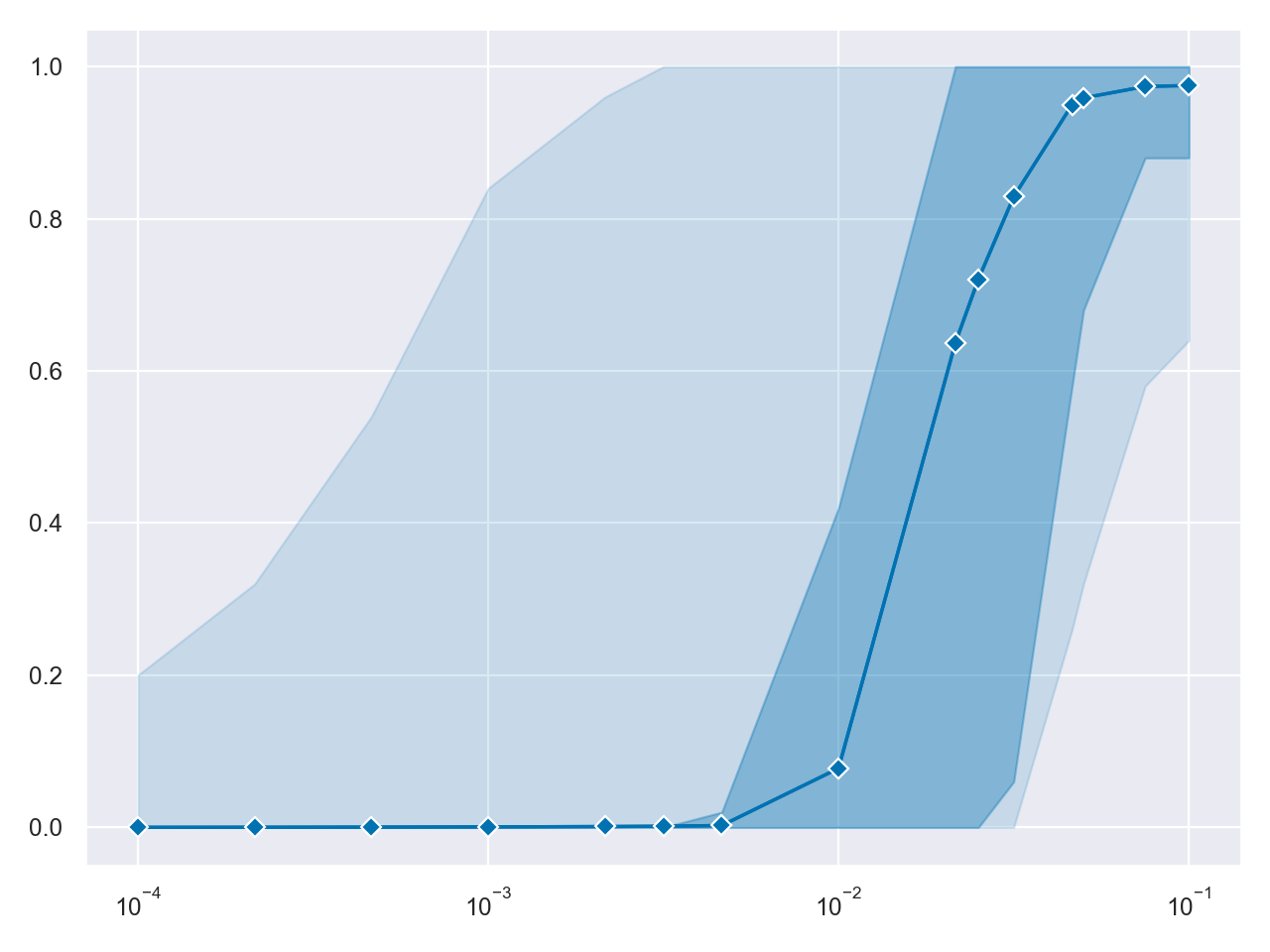}
        \caption{21,18}
    \end{subfigure}
    \hfill
    \begin{subfigure}{0.45\linewidth}
        \centering
        \includegraphics[width=\linewidth]{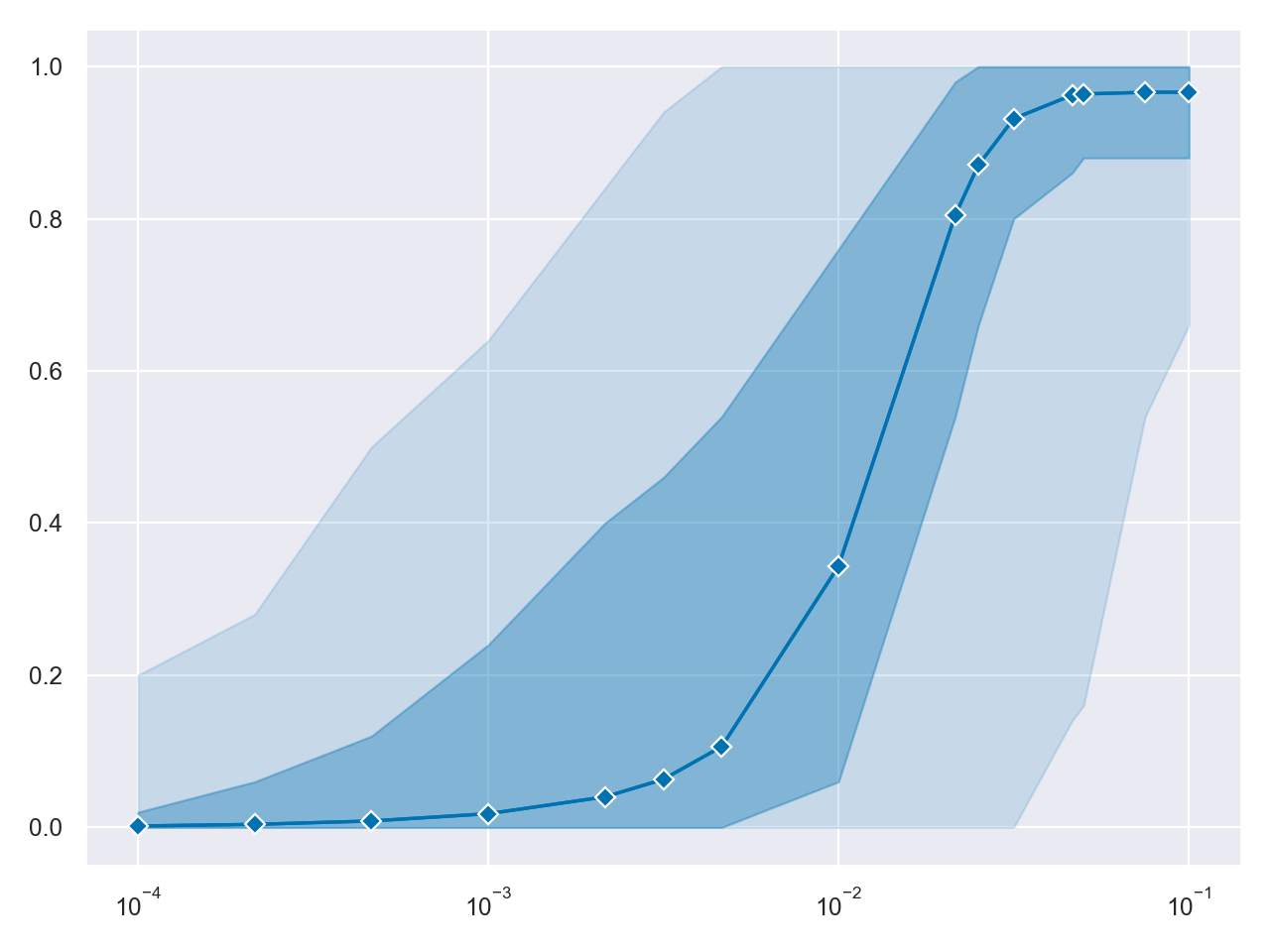}
        \caption{3,27}
    \end{subfigure}
        \vfill
    \begin{subfigure}{0.45\linewidth}
        \centering
        \includegraphics[width=\linewidth]{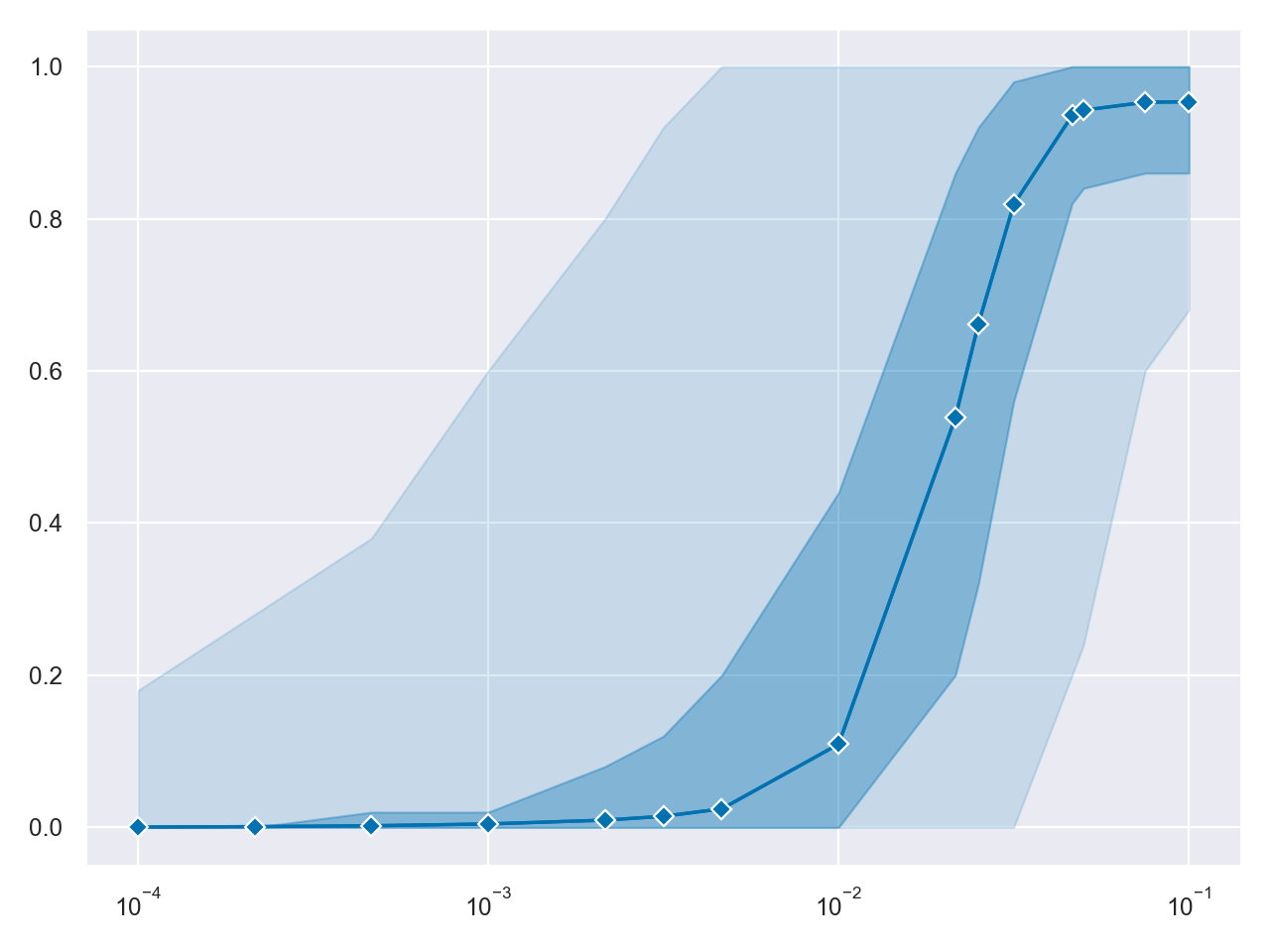}
        \caption{5,15}
    \end{subfigure}
    \hfill
    \begin{subfigure}{0.45\linewidth}
        \centering
        \includegraphics[width=\linewidth]{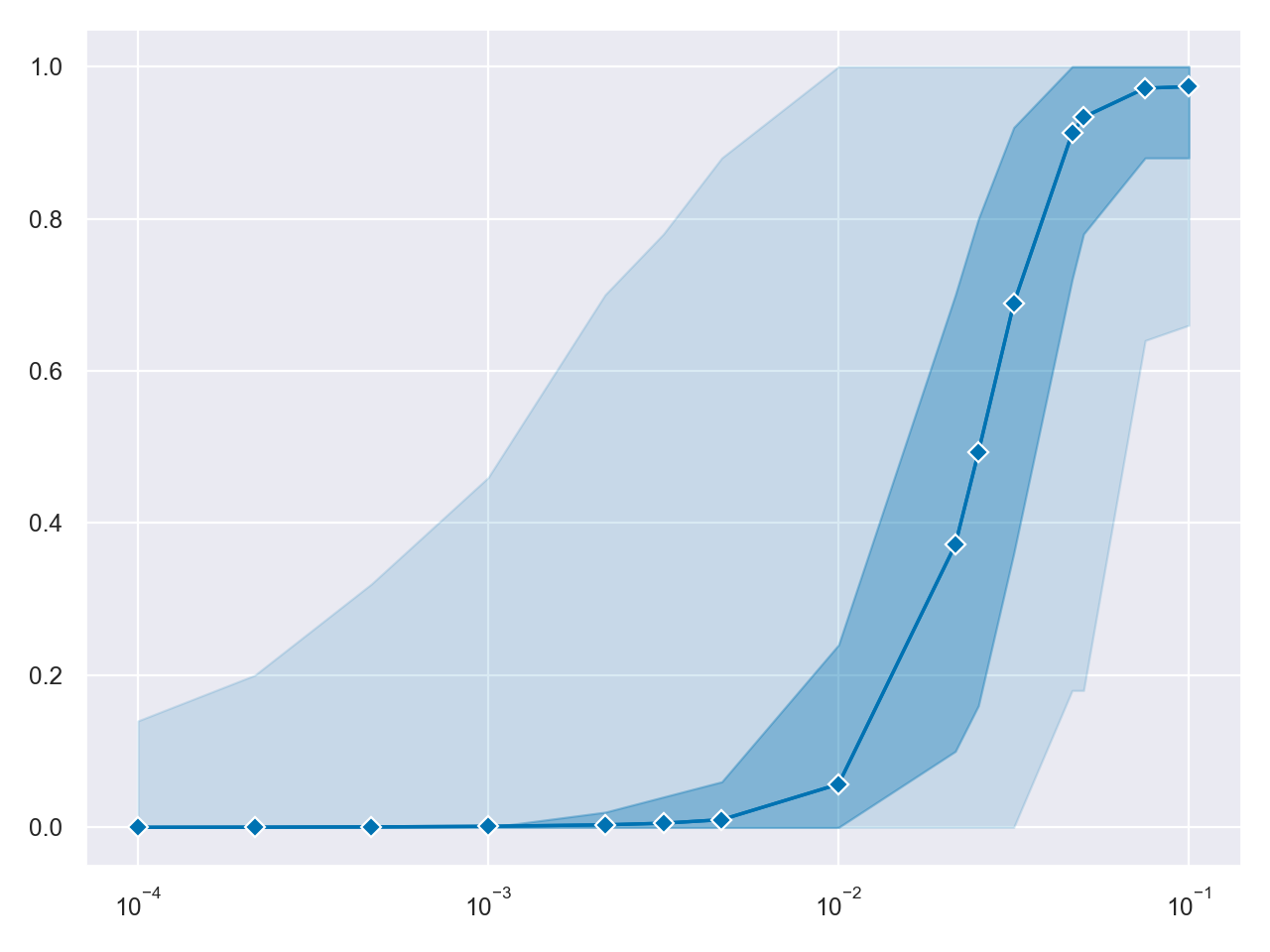}
        \caption{9,6}
    \end{subfigure}
    \caption{Physical error rate to logical error rate with \(100\%\) variance.}
\end{figure*}

\begin{figure*}[htbp]
    \centering
    \begin{subfigure}{0.45\linewidth}
        \centering
        \includegraphics[width=\linewidth]{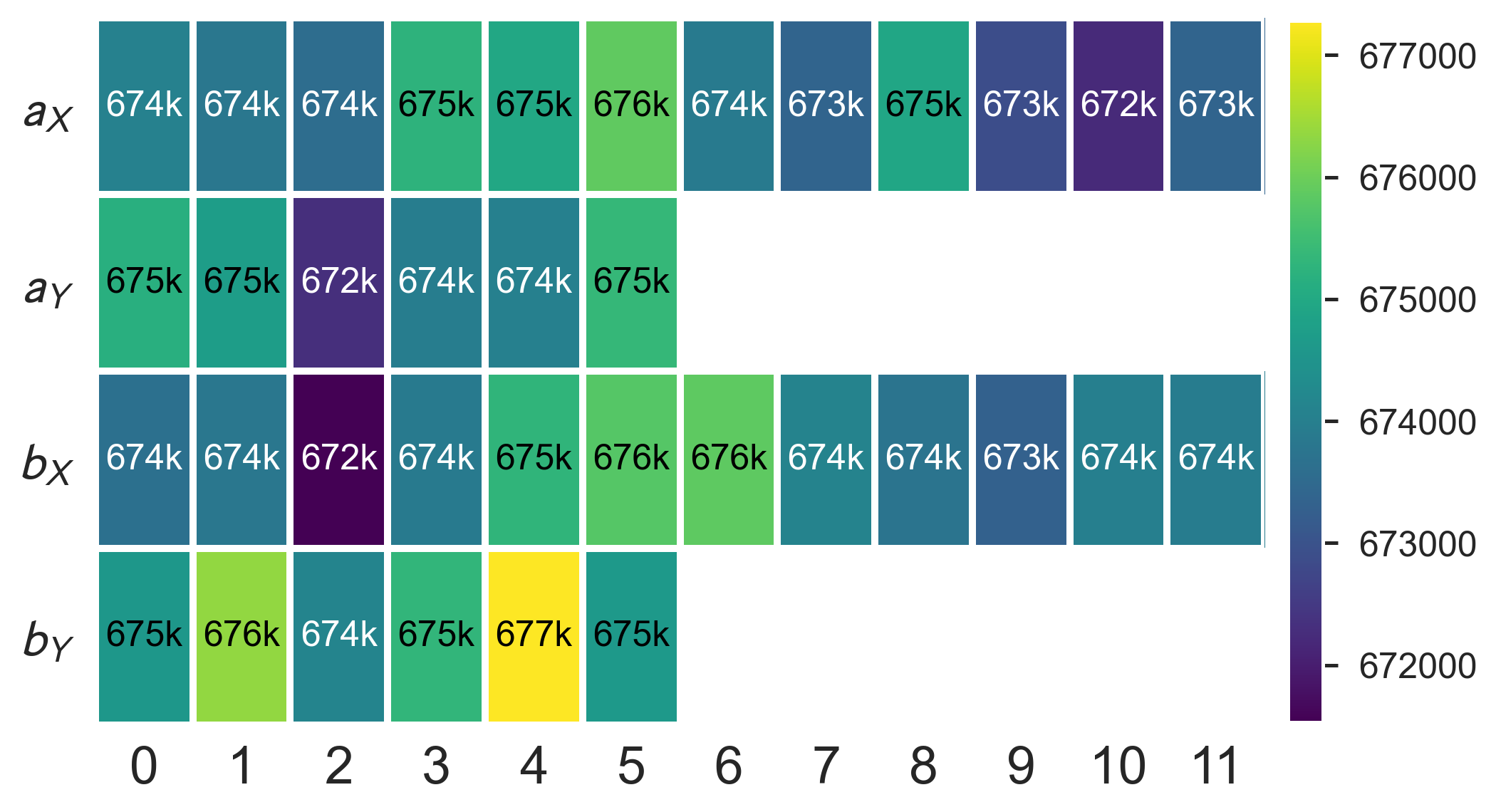}
        \caption{12,6}
    \end{subfigure}
    \hfill
    \begin{subfigure}{0.45\linewidth}
        \centering
        \includegraphics[width=\linewidth]{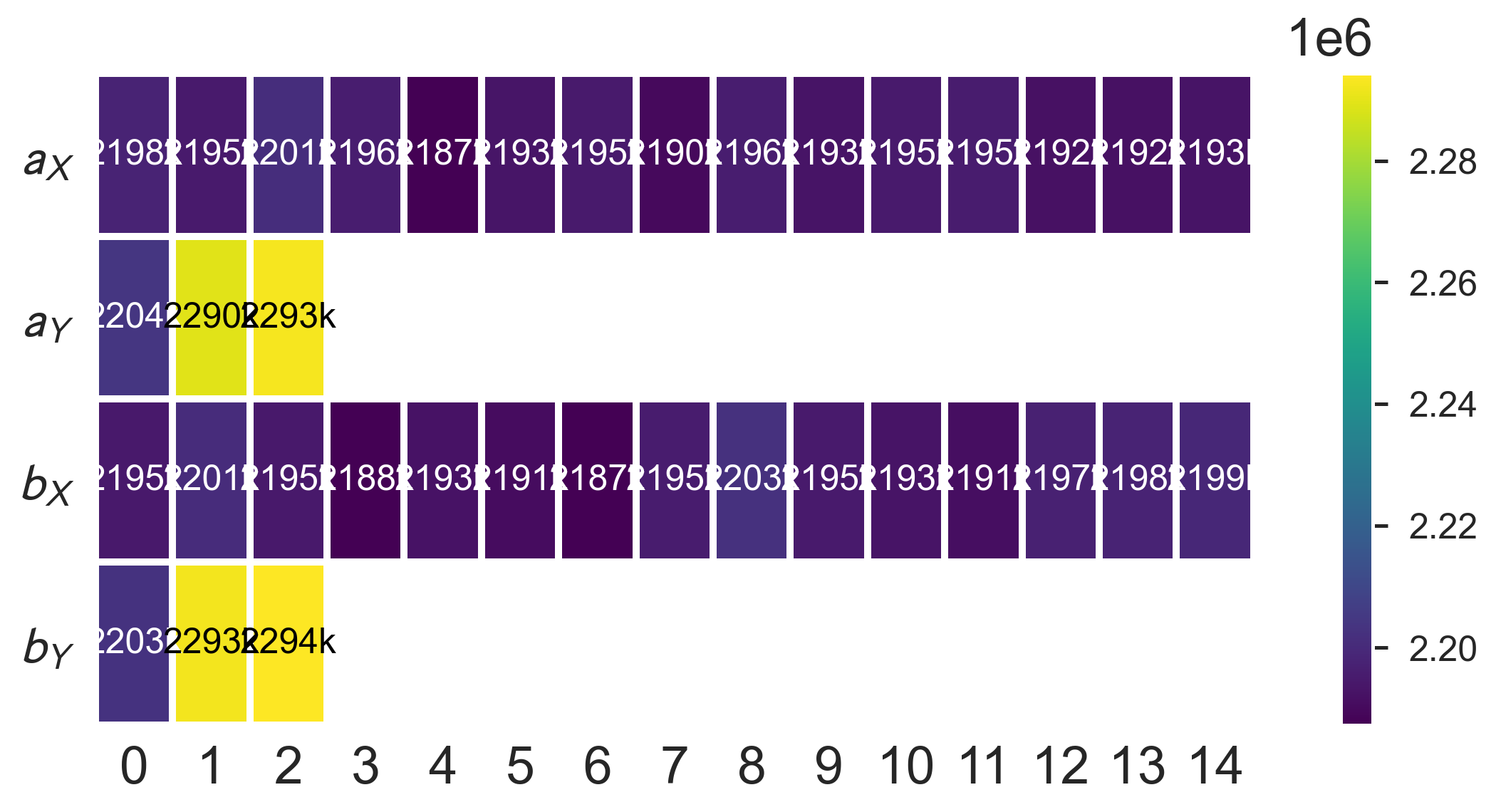}
        \caption{15,3}
    \end{subfigure}
    \vfill
    \centering
    \begin{subfigure}{0.45\linewidth}
        \centering
        \includegraphics[width=\linewidth]{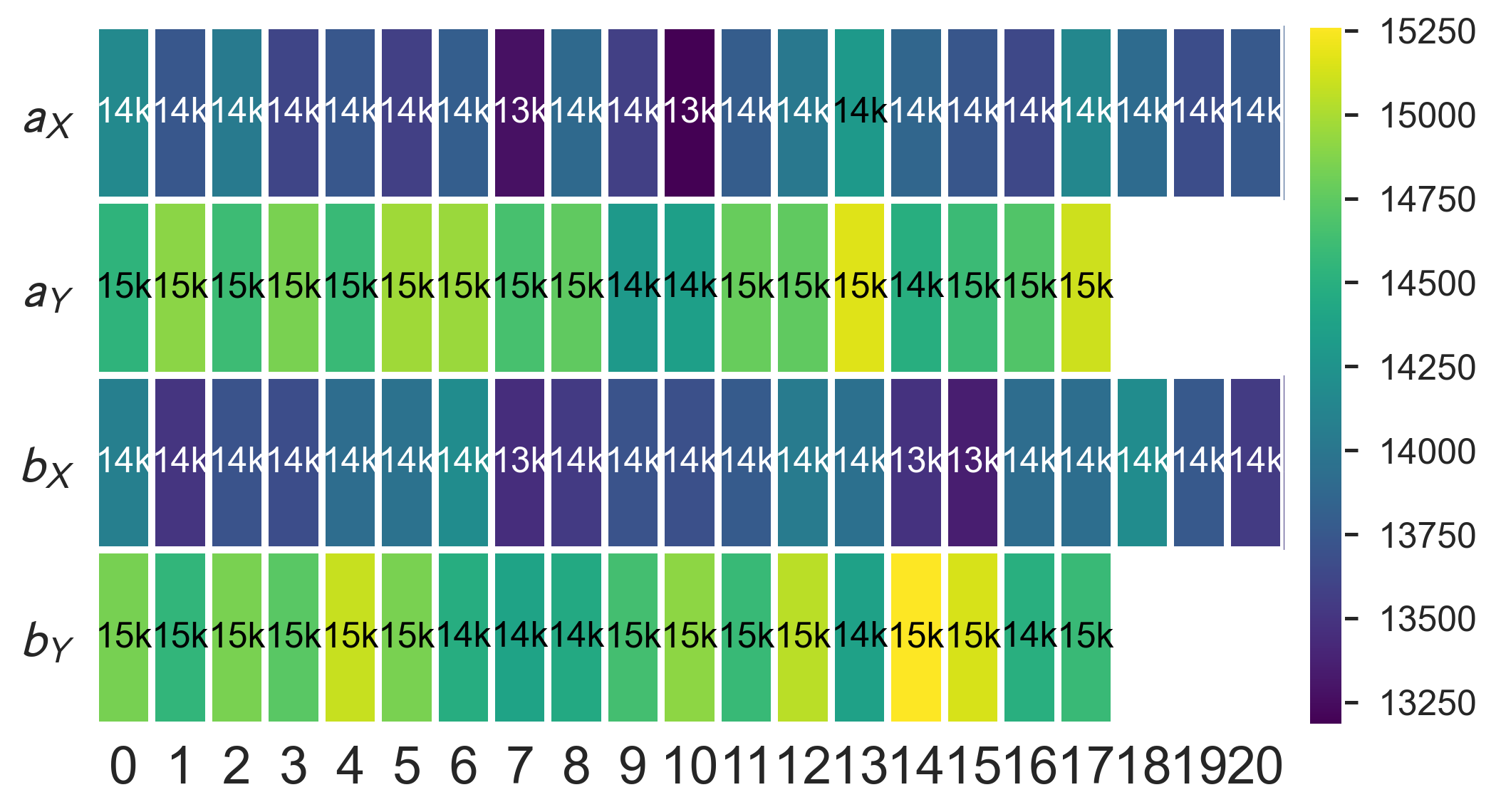}
        \caption{21,18}
    \end{subfigure}
    \hfill
    \begin{subfigure}{0.45\linewidth}
        \centering
        \includegraphics[width=\linewidth]{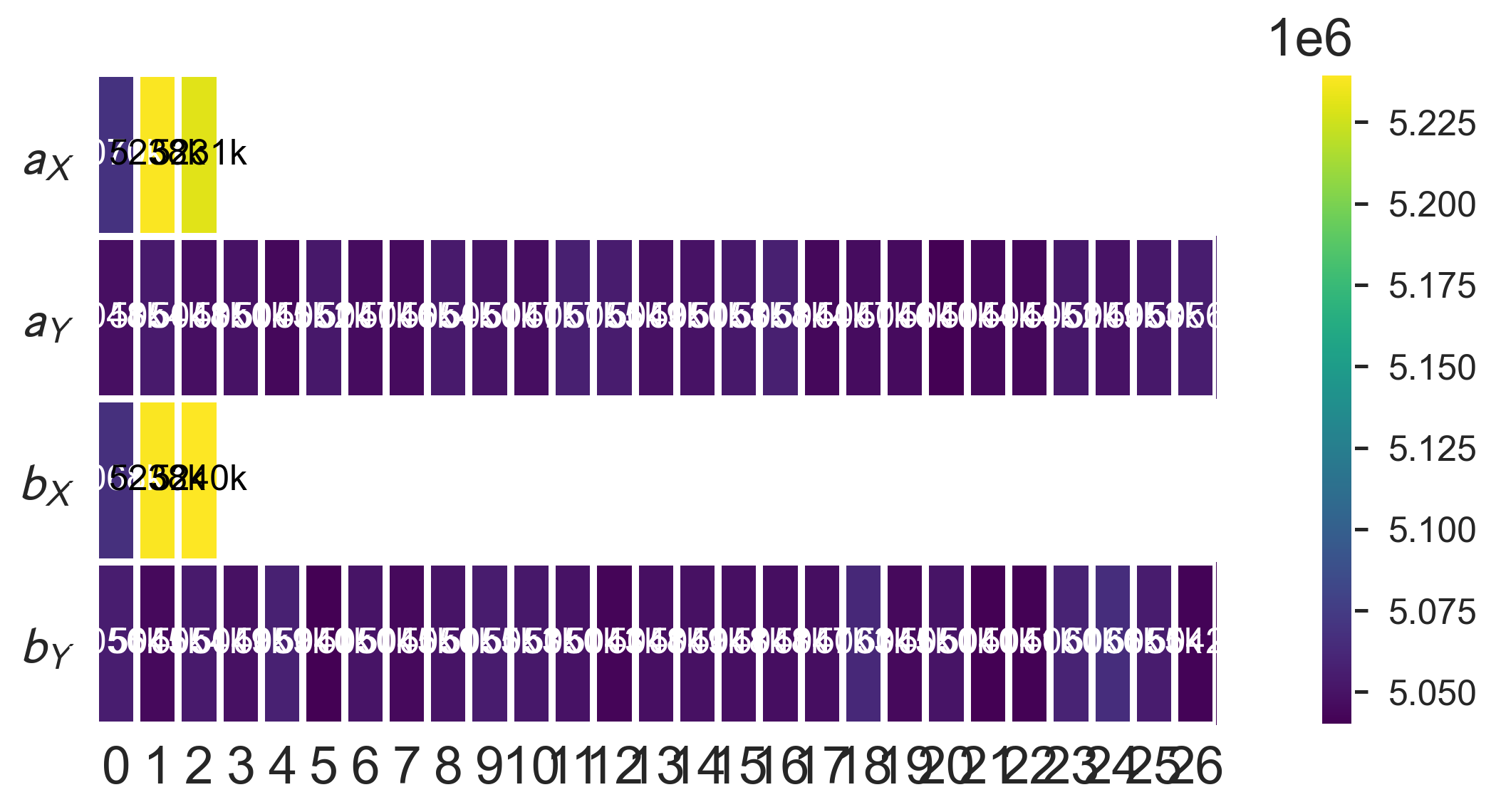}
        \caption{3,27}
    \end{subfigure}
    \vfill
    \centering
    \begin{subfigure}{0.45\linewidth}
        \centering
        \includegraphics[width=\linewidth]{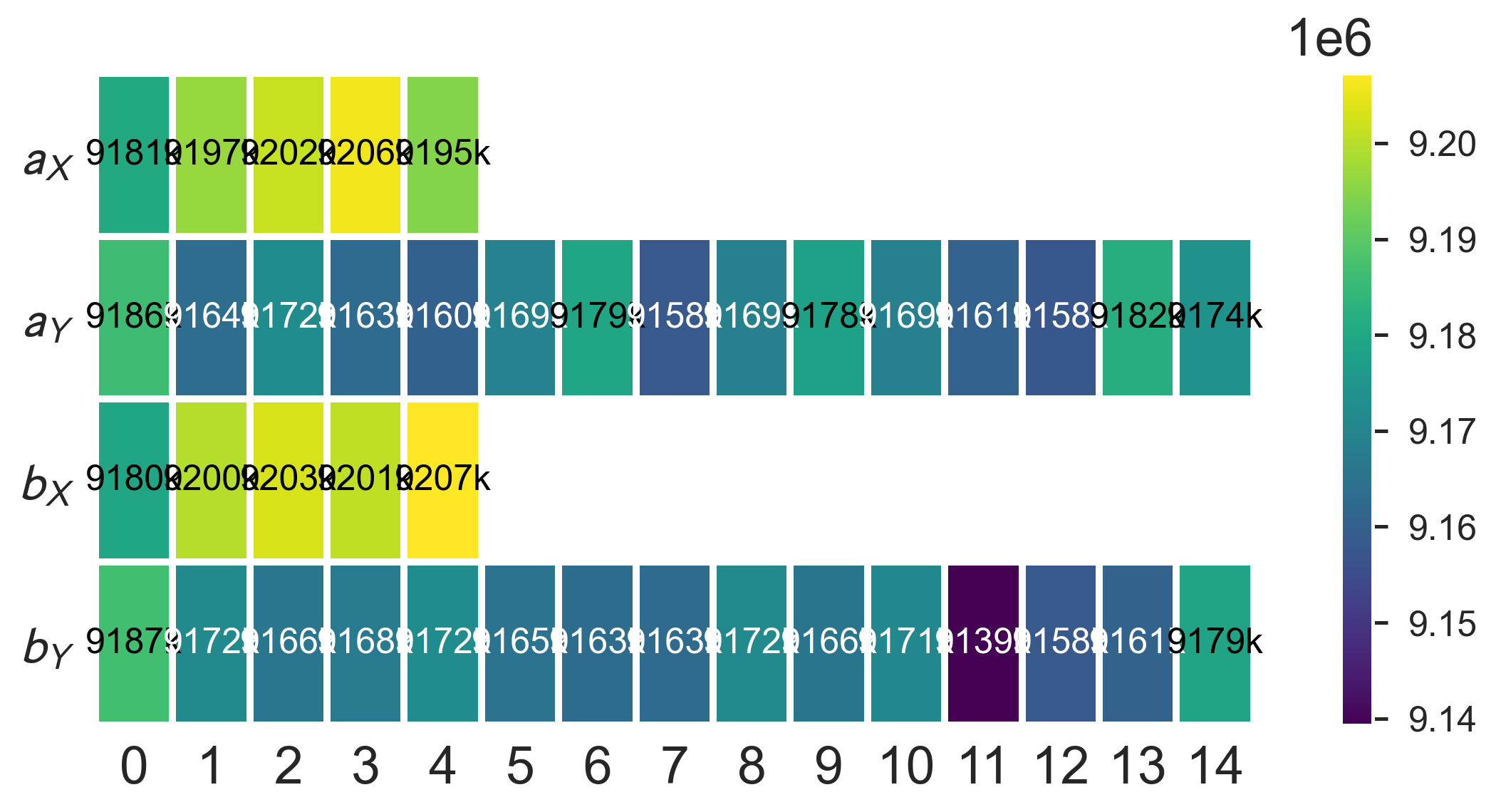}
        \caption{5,15}
    \end{subfigure}
      \hfill
    \begin{subfigure}{0.45\linewidth}
        \centering
        \includegraphics[width=\linewidth]{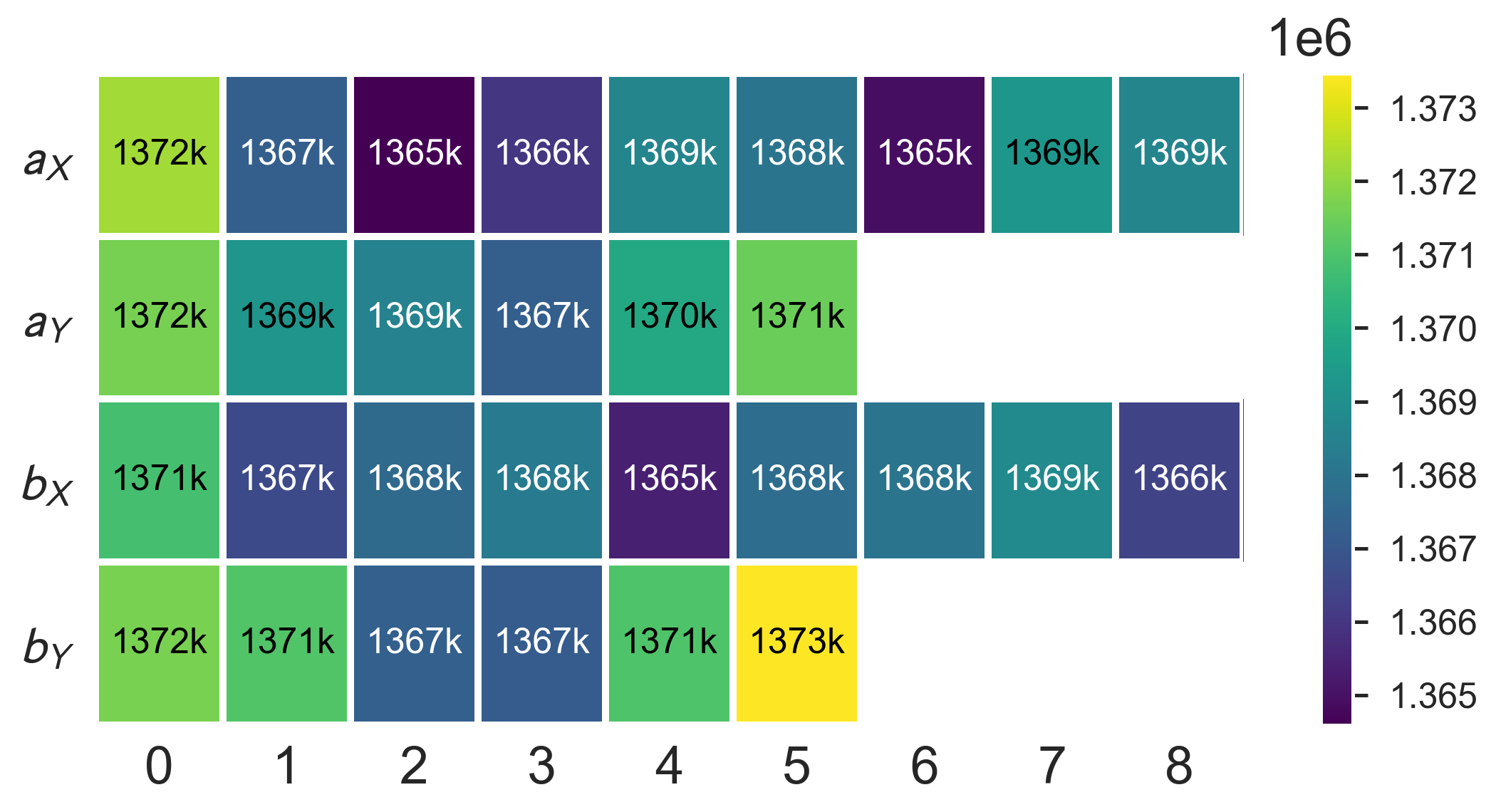}
        \caption{9,6}
    \end{subfigure}
    \caption{Coefficient heat map.}
    \label{fig:coefficientsHeatmap-rest}
\end{figure*}

\bibliography{bibliography}

\end{document}